\documentclass[11pt]{article}
\usepackage[margin=1in]{geometry}
\usepackage[T1]{fontenc}
\usepackage{lmodern}
\usepackage{amsmath,amssymb,amsthm,mathtools}
\usepackage{thm-restate}
\makeatletter
\@ifundefined{newcounteralias}{}{%
  \renewcommand\thmt@autorefsetup{%
    \@xa\def\csname\thmt@envname autorefname\@xa\endcsname\@xa{\thmt@thmname}}%
}
\makeatother
\usepackage{microtype,booktabs,array,enumitem}
\usepackage{needspace,flafter}
\usepackage[dvipsnames]{xcolor}
\usepackage{tcolorbox}
\usepackage{algorithm}
\usepackage{algpseudocode}
\usepackage[colorlinks=true,linkcolor=MidnightBlue,citecolor=MidnightBlue,urlcolor=MidnightBlue]{hyperref}
\usepackage[nameinlink,capitalize,noabbrev]{cleveref}
\newcounter{procedure}
\crefname{procedure}{Procedure}{Procedures}
\Crefname{procedure}{Procedure}{Procedures}
\makeatletter
\providecommand*{\theHALG@line}{}
\renewcommand*{\theHALG@line}{\thealgorithm.\arabic{ALG@line}}
\makeatother
\newtheorem{theorem}{Theorem}
\newtheorem{lemma}[theorem]{Lemma}
\newtheorem{proposition}[theorem]{Proposition}
\newtheorem{corollary}[theorem]{Corollary}
\theoremstyle{definition}
\newtheorem{definition}[theorem]{Definition}
\newtheorem{remark}[theorem]{Remark}
\newcommand{\R}{\mathbb R}
\newcommand{\N}{\mathbb N}
\newcommand{\E}{\mathbb E}
\newcommand{\Prb}{\mathbf{Pr}}
\newcommand{\Var}{\operatorname{Var}}
\newcommand{\KL}{\operatorname{KL}}
\newcommand{\TV}{\operatorname{TV}}
\newcommand{\Lap}{\operatorname{Lap}}
\newcommand{\abs}[1]{\left\lvert#1\right\rvert}
\newcommand{\Unif}{\operatorname{Unif}}
\newcommand{\err}{\operatorname{err}}
\newcommand{\poly}{\operatorname{poly}}
\newcommand{\norm}[1]{\left\lVert#1\right\rVert}
\newcommand{\ip}[2]{\left\langle#1,#2\right\rangle}
\newcommand{\eps}{\varepsilon}
\newcommand{\cQ}{\mathcal Q}
\newcommand{\cY}{\mathcal Y}
\newcommand{\cS}{\mathcal S}

\numberwithin{equation}{section}
\allowdisplaybreaks[2]
\hypersetup{pdftitle={Efficient Pure-DP Statistical Query Release via Sparse Input Smoothing}}
\title{\textbf{Pure Differentially Private Query Release\\
via Laplace Input Smoothing}}
\author{Zongrui Zou\thanks{State Key Laboratory for Novel Software Technology, Nanjing University (\texttt{zou.zongrui@smail.nju.edu.cn}).}}
\date{}
\begin{document}
\maketitle

\begin{abstract}
We study the offline release of a public family of \(k\) statistical
queries on a database of \(N\) records from a public universe of size
\(d\), under pure $\eps$-differential privacy.  We give an \emph{efficient} algorithm
that, with probability at least \(1-\beta\), answers all queries
simultaneously to error
\[
 O\!\left(\min\left\{1,
       \sqrt{\frac{\log(d/\beta)\log k}{N\eps}}\right\}\right).
\]
The output is a synthetic histogram on the data universe, and its running time depends polynomially on $N,d$ and $k$.

This resolves a conjecture of Nikolov and Ullman (2021) with an efficient, polynomial-time construction. Fitzsimons~\cite{Fitzsimons26} recently established the same square-root error scale via an information-theoretic construction that relies on a computationally intractable Lipschitz envelope over the exponential space of possible outputs. In contrast, we introduce an input-side smoothing approach: we run a selection-only multiplicative-weights process after a single \textit{sparse} Laplace perturbation of the input histogram. Departing from standard first-order stability analyses, we investigate the \emph{stability of stability} by establishing a uniform lower bound on the second-order log-derivatives of the output transcript across $T$ rounds of multiplicative weight updates. We then develop an analytic smoothing principle showing that this curvature bound allows a sparse, coordinate-wise perturbation to guarantee pure differential privacy globally, attaining optimal utility through a conceptually simple, polynomial-time pure-DP mechanism.
\end{abstract}
\thispagestyle{empty}
\newpage
\setcounter{page}{1}
\section{Introduction}\label{sec:introduction}

Extracting statistical insights from large-scale collections of data serves as a fundamental task of modern data analytics or machine learning. At the heart of this paradigm lies the framework of \emph{statistical queries} 
\cite{kearns1998efficient}: evaluating the empirical average of a bounded predicate 
over an underlying dataset. 
 Formally, consider a dataset $D = (x_1, \dots, x_N) \in \mathcal{X}^N$ comprising 
$N$ records drawn from an arbitrary data universe $\mathcal{X}$. A statistical query 
is specified by a bounded evaluation predicate $\phi \colon \mathcal{X} \to [0, 1]$. 
The exact empirical answer to the query $\phi$ on dataset $D$ is defined as the sample mean:
\begin{equation}\label{eq:sq-def}
  q_\phi(D) \coloneqq \frac{1}{N} \sum_{i=1}^N \phi(x_i) \in [0, 1].
\end{equation}


Such queries encompass a vast spectrum of empirical measurements, ranging from 
demographic frequency tables and low-order marginals in census surveys 
\cite{barak2007privacy,abowd2018us,LT25,ZLU25} to empirical 
risk minimization (ERM), loss function evaluation, and gradient computation in 
statistical learning \cite{kasiviswanathan2011can,chaudhuri2011differentially,bassily2014private,bassily2019private,feldman2020private,lowy2023private}.

However, the utility from answering extensive query workloads comes hand in hand with severe privacy risks. Even when each query averages over 
thousands of individuals, releasing a large collection of accurate statistics can inadvertently leak sensitive personal 
attributes~\cite{dinur2003revealing}, exposing sensitive medical records, financial transactions, or browsing telemetry. To mitigate the risk of exposing personal records, \emph{differential privacy} (DP) 
\cite{DMNS06,DR14} has been established as the gold 
standard for privacy-preserving data analysis. Differential privacy 
guarantees that a randomized algorithm's output distribution remains 
indistinguishable whether any single individual's record is replaced. Formally,

\begin{definition}[Differential Privacy \cite{DMNS06}]\label{def:dp-intro}
Let $\mathcal{X}$ be an arbitrary data domain, and let $\mathcal{M}$ be a randomized 
algorithm mapping databases to an output space $\mathcal{Y}$. For privacy parameters 
$\eps > 0$ and $\delta \in [0, 1)$, we say that $\mathcal{M}$ satisfies 
$(\eps, \delta)$-differential privacy if, for every pair of neighboring databases 
$D, D'$ differing by the replacement of at most one individual record, and for every 
measurable set of outcomes $A \subseteq \mathcal{Y}$,
\[
  \Prb[\mathcal{M}(D) \in A] \le e^\eps \Prb[\mathcal{M}(D') \in A] + \delta.
\]
When $\delta = 0$, we say that $\mathcal{M}$ preserves \emph{pure} differential 
privacy (denoted as pure $\eps$-DP).
\end{definition}

In this paper, we study privately releasing 
an arbitrary workload of statistical queries under pure differential privacy in an offline manner. 
Formally, let $[d] = \{1, \dots, d\}$ be a fixed public data universe, and consider 
a database of $N$ records drawn from $[d]$. Such a database is 
uniquely represented by its histogram $h \in \N^d$ with $\|h\|_1 = \sum_{j=1}^d h_j = N$, where $h_j$ 
denotes the number of individuals with attribute $j$. We denote the feasible set of all such valid 
histograms by $\mathcal{H}_{N,d}$. Under record replacement, two databases are 
neighboring if one record is modified; in histogram space, 
this corresponds to the adjacency $h' = h + e_a - e_b$ for some 
$a, b \in [d]$, where $e_j \in \R^d$ is the $j$-th standard basis vector. Note that 
$\|h - h'\|_1 \le 2$. The workload consists of a public family of $k$ statistical 
queries $\cQ = \{q_1, \dots, q_k\}$, where each query is a bounded function 
$q_i: [d] \to [0, 1]$. For an underlying database $h$, according to the definition in~\eqref{eq:sq-def}, the true answer to 
query $q_i$ is the linear average
\[
  q_i(h) \coloneqq \frac{\ip{h}{q_i}}{N} = \frac{1}{N}\sum_{j=1}^d h_j q_i(j) \in [0, 1].
\]
Our objective is to design a private algorithm that on the input histogram $h$, releases an empirical answer vector $\widehat a = (\widehat a_1, \dots, \widehat a_k) \in \R^k$ 
(or more desirably, a public synthetic distribution $\widehat p \in \Delta_d$ with 
answers $\widehat a_i = \ip{\widehat p}{q_i}$) that minimizes the simultaneous 
worst-case $\ell_\infty$ error across the entire workload:
\begin{equation}\label{eq:minimize-goal}
  \err_{\cQ}(\widehat a, h) \coloneqq \max_{i \in [k]} \left| \widehat a_i - q_i(h) \right|
  = \max_{i \in [k]} \left| \widehat a_i - \frac{\ip{h}{q_i}}{N} \right|.
\end{equation}

Privately releasing statistical queries stands as a cornerstone problem in differential privacy, having inspired a rich body of work over the past two decades~\cite{DNRRV09,RR10,HR10,DRV10,HLM12,GRU12,BLR13,ENU20,Nikolov23,NT24,LT25,ZLU25,Fitzsimons26}. Releasing a \emph{single} statistical query under differential privacy is straightforward: because replacing one individual's record can change the empirical 
average $q_i(h)$ by at most $1/N$, the $\ell_1$-sensitivity of any single query is $1/N$. 
Calibrating independent noise from the Laplace mechanism \cite{DMNS06} therefore incurs 
an expected error of only $O(1/(N\eps))$. However, releasing an entire workload of $k$ 
queries introduces a scalability challenge. Naively adding independent noise 
to each query causes the maximum error across the workload to grow as 
$O(k\log(k/\beta)/(N\eps))$, which becomes useless as soon as the workload 
size $k$ exceeds the sample size $N$. Yet, these $k$ answers are far from independent: 
they represent correlated linear evaluations over the exact same underlying distribution 
on $[d]$. A better private mechanism must exploit this shared statistical structure 
to decouple the accuracy from the query count.

The feasibility of bypassing the polynomial dependence on $k$ was first established by Blum, Ligett, and Roth \cite{BLR13}. They observed that for any 
finite workload $\cQ$ of size $k$, a small synthetic sub-database consisting of only 
$s = O(\log(2k)/\alpha^2)$ records uniformly sampled from the whole dataset can 
approximate all $k$ queries to within additive error $\alpha$. Because there are at most 
$d^s$ such ordered candidate sub-databases, selecting an accurate candidate via the 
exponential mechanism \cite{MT07} incurs a selection error of $O(s\log d / (N\eps))$. 
Balancing the approximation error $\alpha$ against this selection penalty yields an overall 
error of
\[
  \alpha + \frac{\log(2k)\log d}{N\eps\,\alpha^2} 
  \quad\Longrightarrow\quad 
  O\!\left( \left(\frac{\log(2d)\log(2k)}{N\eps}\right)^{1/3} \right).
\]
While this gives a logarithmic dependency on the query count $k$, it is not efficiently computable: the candidate space has size $d^{O(\log(2k)/\alpha^2)}$, 
making an exhaustive search slow. Soon after, Dwork, Naor, Reingold, 
Rothblum, and Vadhan \cite{DNRRV09} showed that computational efficiency is fundamentally 
distinct from information-theoretic existence. They established that under standard 
cryptographic hardness assumptions, any computationally efficient mechanism must explicitly tolerate 
a polynomial dependence on the data universe size $d$ in its running time. In light of this computational barrier, the primary objective for follow-up work is to design a private mechanism whose running time is polynomial in the database size, the universe size, and the query count, namely $\operatorname{poly}(N, d, k)$.

To achieve computational efficiency, a sequence of pioneering works shifted the paradigm from brute-force selection over candidate synthetic representations to iterative, interactive data release.
Roth and Roughgarden \cite{RR10} introduced the median mechanism, maintaining a version 
space of plausible databases that is iteratively pruned whenever an informative query 
reveals a significant discrepancy. Hardt and Rothblum \cite{HR10} replaced the combinatorial 
version space with a continuous distribution over $[d]$ updated via multiplicative weights 
(the Private Multiplicative Weights, or PMW, framework), guaranteeing running times 
polynomial in $d$. Concurrently, Dwork, Rothblum, and Vadhan \cite{DRV10} developed a 
boosting-based perspective, showing that weak synopses can be iteratively improved to achieve 
polylogarithmic query dependence. This iterative paradigm was subsequently unified by 
Gupta, Roth, and Ullman \cite{GRU12} under the abstraction of iterative database 
construction: in each round, a private selection rule identifies an informative distinguishing 
query on which the current model fails, and an optimization update makes monotonic progress.

Remarkably, this iterative framework revealed a persistent algorithmic gap 
between \emph{approximate} $(\eps, \delta)$-differential privacy and \emph{pure} $\eps$-differential 
privacy. Under approximate differential privacy, advanced composition theorems \cite{DRV10,DR14} 
guarantee that the privacy loss of $T$ adaptive steps concentrates sub-linearly around 
$\sqrt{T}\eps_0$. This concentration allows one to allocate a generous per-round privacy 
budget $\eps_0 \asymp \eps/\sqrt{T}$, which drives the overall error down to the optimal 
$N^{-1/2}$ scaling, yielding errors of the form $\widetilde{O}((\log d)^{1/4}\sqrt{\log k} / \sqrt{N\eps})$ 
\cite{HR10,GRU12,LT25}. Under pure differential privacy ($\delta = 0$) where advanced composition is not applicable, one 
is constrained by linear basic composition: the cumulative privacy loss of $T$ rounds is 
simply $\sum_{t=1}^T \eps_t = T\eps_0$. 

The algorithmic consequence and the long-standing error gap caused by this difference in privacy accounting between pure and approximate DP are illustrated most transparently by the Multiplicative Weights Exponential Mechanism (MWEM) of Hardt, Ligett, and McSherry~\cite{HLM12}. In each of $T$ iterations, 
MWEM uses the exponential mechanism to select an inaccurate query and adds independent Laplace 
noise to measure its true value, updating a candidate distribution over $[d]$. The potential 
argument of multiplicative weights guarantees that the empirical regret decays as 
$O(\sqrt{(\log d)/T})$. Simultaneously, under pure-DP, 
each of the $T$ rounds is allocated only $\eps/T$ privacy, meaning the noise added to the 
measurements and selection scores must scale as $O(T\log(2k)/(N\eps))$. Combining these two 
opposing forces yields the MWEM error balance:
\begin{equation}\label{eq:intro-old-balance}
  \err_{\cQ} \approx O\!\left( \sqrt{\frac{\log d}{T}} + \frac{T\log(2k)}{N\eps} \right).
\end{equation}
Therefore, while balancing this trade-off under pure privacy, the second term grows linearly with $T$. 
Optimizing $T$ by setting $T \asymp (N\eps)^{2/3}$, gives the overall error in the cube-root rate 
$\widetilde{O}(N^{-1/3})$ \cite{HLM12}. By contrast, 
if approximate privacy with advanced composition were used, the second term in 
\eqref{eq:intro-old-balance} would instead scale as $\frac{\sqrt{T}\log(2k)}{N\eps}$, 
which balances against $\sqrt{(\log d)/T}$ at $T \asymp N\eps$ to achieve 
the square-root scaling $\widetilde{O}(N^{-1/2})$. For over a decade, this disparity 
raised the question of whether the cube-root scaling is an inescapable price of pure privacy 
in iterative mechanisms, or merely a bottleneck of worst-case step-by-step composition.

Geometric lower bounds strongly indicated that this cube-root rate may not be fundamental. 
Using packing arguments, Hardt and Talwar \cite{HT10} and 
Hardt \cite{Hardt11} established that the worst-case information-theoretic lower bound 
for pure-DP query release scales as $\widetilde{\Omega}(N^{-1/2})$, mirroring the approximate-DP 
fingerprinting lower bounds of Bun, Ullman, and Vadhan \cite{BUV14} and Lyu and Talwar 
\cite{LT25}. Recognizing this gap, Nikolov and Ullman \cite{NU21} formally conjectured that 
the true worst-case maximum error for pure-DP statistical query release satisfies
\begin{equation}\label{eq:intro-target}
  O\!\left( \sqrt{\frac{\log k \log d}{N\eps}} \right).
\end{equation}
While progress was made on related geometric objectives, such as instance-optimal bounds 
for unconstrained linear queries via the matrix mechanism \cite{LMHMR15,NTZ13,ENU20,NT24,ZLU25} 
and dimension-reduced $\ell_2$ average-error guarantees \cite{Nikolov23}, the maximum error 
$\ell_\infty$ conjecture of Nikolov and Ullman remained unresolved for general workloads. 
Recent purification frameworks based on output post-processing \cite{LWMW25} still incurred 
an accuracy overhead that preserved the $N^{-1/3}$ cube-root barrier.

Very recently, Fitzsimons~\cite{Fitzsimons26} resolved the 
\emph{existence} aspect of the Nikolov--Ullman conjecture in the affirmative. Fitzsimons considered an idealized, \emph{selection-only} multiplicative weights process: 
in each of $T$ rounds, rather than measuring noisy query answers, the algorithm merely 
privately selects an inaccurate query among the $2k$ signed directions $\{\pm q_1, \dots, \pm q_k\}$ 
(accounting for both over- and under-estimation), producing a transcript of selected 
indices $y = (i_1, \dots, i_T) \in [2k]^T$ with output distribution $p_D(y)$ on input $D$. To privatize this process under pure DP without paying step-by-step composition penalties, 
Fitzsimons~\cite{Fitzsimons26} information-theoretically reweighted each output probability according to its Lipschitz upper 
envelope over all candidate databases:
\[
  \sup_{D'} e^{-(\eps/2)\operatorname{dist}(D,D')} p_{D'}(y),
\]
where $\operatorname{dist}(D,D')$ counts record replacements between databases. 
By showing that this envelope distribution on outputs $y$ remains close to the idealized, unperturbed MW process in R\'enyi divergence, Fitzsimons proved that the expected error achieves the conjectured square-root rate 
\eqref{eq:intro-target}. However, this construction is purely an 
existence proof: evaluating a single output's probability requires taking a supremum 
over all possible databases $D'$, and sampling requires normalizing over the exponentially 
large transcript space $[2k]^T$. Because both operations are computationally intractable, 
the challenge of designing a polynomial-time algorithm remained entirely open. 

\textbf{Our Contribution.} We give a polynomial-time pure-private mechanism that attains the $N^{-1/2}$ rate in $N$ and a $\operatorname{polylog}(d,k)$ rate in $d$ and $k$ for statistical query release. This confirms the Nikolov--Ullman conjecture (stated in \eqref{eq:intro-target}) via a polynomial-time computable algorithm.

To achieve efficiency, we adopt a \emph{different} perspective compared to Fitzsimons~\cite{Fitzsimons26}. Rather than attempting to repair an exponentially large output distribution, we take the dual approach: We preprocess the input histogram $h \in \mathbb{R}^d$ by adding very sparse Laplace perturbation before running the selection process. Here, we add Laplace perturbation \emph{not} to privatize the histogram, but to \emph{smooth} $p_h(y)$, the distribution of transcripts $y$ later produced by the iterative selection.

Specifically, in our iterative process, \emph{bad} queries (those on which the current model still fails, so that we can update the model accordingly) are drawn via exponential mechanism with parameter $\lambda$. For any fixed $T$-round transcript $y$, differentiating its log-probability twice with respect to an input coordinate $h_j$ ensures a uniform lower bound $\partial_{h_j}^2 \log p_h(y) \ge -T\lambda^2$. 
We establish an analytic smoothing principle showing that this log-curvature bound prevents likelihood ratios from collapsing too rapidly, allowing a sparse input perturbation (e.g., adding Laplace noise to each coordinate with a tiny probability $\pi = \Theta(\beta/d)$ and leaving it untouched otherwise) to globally bound the multiplicative privacy loss. By combining 
this input perturbation with an unperturbed martingale analysis of selection-only MW, 
our mechanism avoids the step-by-step composition penalty of MWEM for pure-DP, and achieves the optimal 
square-root rate in $O(dk\lceil N\eps/\log(2k)\rceil)$ time.

\subsection{Our Results}

Our algorithm releases a weighted histogram, represented by a probability distribution \(\widehat p\in\Delta_d\).\footnote{$\Delta_d = \{x\in \mathbb{R}^d| \sum x_i= 1 \text{ and } x_i \geq 0 \text{ for all } i\in[d]\}$.}
It answers query \(q_i\) by \(\ip{\widehat p}{q_i}\).
For any released distribution \(p\in\Delta_d\), we consider minimizing its $\ell_\infty$ query error defined as
\[
 \err_{\cQ}(p,h)
 :=\max_{i\in[k]}
       \left|\ip{p}{q_i}-\frac{\ip{h}{q_i}}N\right|,
\]
which matches \eqref{eq:minimize-goal}.
\begin{theorem}\label{thm:hp}
There is a universal constant \(C\) with the following property.
Let \(N\geq1\), \(d\geq2\), \(k\geq1\), and
\(0<\eps\leq1\), \(0<\beta<1/2\). There is a pure \(\eps\)-differentially private algorithm (under replacement adjacency) such that on every histogram \(h\) of mass \(N\) and every public query set \(\{q_i\}_{i\in [k]}\in[0,1]^{k\times d}\), it outputs \(\widehat p\in\Delta_d\) and
satisfies
\[
 \Prb\!\left\{\err_{\cQ}(\widehat p,h)>
 C\min\left\{1,
       \sqrt{\frac{\log(2d/\beta)\log(2k)}{N\eps}}\right\}\right\}
 \leq\beta .
\]
Further, the running time is in $\poly(d,k,N)$.
\end{theorem}

At constant \(\beta\), the displayed scale is already achievable
information-theoretically by applying Markov's inequality to the
expected-error guarantee of~\cite{Fitzsimons26}.
Our result supplies a polynomial-time construction under the
computational assumptions of \cref{rem:computation}, and with a weighted synthetic output.
Note that, from the perspective of sample complexity, error at most \(\alpha\) is achieved with
\(N=O(\log(2d/\beta)\log(2k)/(\eps\alpha^2))\) records. Compared with the previous best efficient pure-DP algorithm for statistical query release~\cite{HLM12}, the dependence on the target accuracy improves from \(\alpha^{-3}\) to \(\alpha^{-2}\).

\paragraph{The remaining gap between pure and approximate privacy.}
Fix a constant failure probability and compare the many-query
scales, keeping the privacy parameter \(\delta\) explicit.  The standard
approximate-private upper bound~\cite{HR10, GRU12} (with matching lower bound~\cite{LT25}) and our pure-private bound respectively require
\begin{equation}\label{eq:pure-approx-compare}
     N_{\mathrm{approx}}
 =O\!\left(\frac{\sqrt{\log d\log(1/\delta)}\log k}
                    {\eps\alpha^2}\right),\qquad
 N_{\mathrm{pure}}
 =O\!\left(\frac{\log d\log k}{\eps\alpha^2}\right).
\end{equation}
Thus a square-root error rate in $N$ does not eliminate the privacy gap: since we usually choose $\delta = o(1/N)$, in the large-dimension regime where $d\gg N$ there can be $\log(1/\delta)\ll\log d$, so approximate-DP algorithms still have better utility than pure-DP algorithms. For small $\delta$ budgets, or in the case where $d = O(N)$ such that $\log(1/\delta)$ is of the same order as $\log d$, the scalings displayed in~\eqref{eq:pure-approx-compare} coincide; for even smaller $\delta$, one may always use a pure-private mechanism.

\begin{table}[h]
\centering
\small
\begin{tabular}{@{}>{\raggedright\arraybackslash}p{.28\linewidth}
 >{\raggedright\arraybackslash}p{.34\linewidth}
 >{\raggedright\arraybackslash}p{.30\linewidth}@{}}
\toprule
Method & Maximum-error scale & Computational efficiency\\
\midrule
Small database~\cite{BLR13}
& \((\log d\log k/(N\eps))^{1/3}\)
& inefficient\\
\addlinespace
Pure MWEM~\cite{HLM12}
& \((\log d\log k/(N\eps))^{1/3}\)
& polynomial in \(N,d,k\)\\
\addlinespace
Transcript envelope~\cite{Fitzsimons26}
& \(\sqrt{\log d\log k/(N\eps)}\)
& inefficient\\
\addlinespace
Input smoothing (this paper)
& \(\sqrt{\log d\log k/(N\eps)}\)
& polynomial in \(N,d,k\)\\
\bottomrule
\end{tabular}
\caption{Comparison of many-query pure-private statistical query release mechanisms.}
\label{tab:comparison}
\end{table}

\section{Technical Overview}\label{sec:overview}
In this section, we give an overview of our main construction and analysis, from the selection-only protocol and the logarithmic derivative view of pure DP to the geometric role of the Hessian and the scalar smoothing lemma. 

\subsection{Selection-Only Multiplicative Weights Update}
\label{sec:overview-framework}

Following the protocol introduced in the recent work of Fitzsimons~\cite{Fitzsimons26}, we use a selection-only interactive learning process and do not \emph{measure} noisy query values at each iteration. Specifically, let $h \in \mathcal{H}_{N, d}$ be the private database histogram, where $h_j \in \N$ counts the multiplicity of attribute $j \in [d]$ and $\|h\|_1= N$. The query family is $\cQ = \{q_1, \dots, q_k\}$, where each $q_i \colon [d] \to [0, 1]$ is represented by a vector $q_i \in [0, 1]^d$. The empirical answer is the normalized inner product $q_i(h) = \ip{h}{q_i}/N$.

The algorithm iteratively updates a candidate synthetic histogram $p\in\Delta_d$ as an approximation of $h/N$. To capture two-sided absolute approximation errors $|\ip{p}{q_i} - \ip{h}{q_i}/N|$ , we define the signed query family $\cS \coloneqq \cQ \cup (-\cQ)$. We index this family as $s_1, \dots, s_{2k}$, where $s_{2i-1} = q_i$ and $s_{2i} = -q_i$ for each $i \in [k]$. For any non-negative candidate histogram $H = Np$, note that its corresponding $\ell_\infty$ query error is:
\begin{equation}\label{eq:overview-signed-max}
  \max_{i \in [2k]} \ip{H - h}{s_i} = N \max_{q \in \cQ} \left| \ip{\frac{H}{N}}{q} - \frac{\ip{h}{q}}{N} \right|.
\end{equation}
Whenever $H$ over-predicts $q$, $\ip{H - h}{q} > 0$; whenever $H$ under-predicts $q$, $\ip{H - h}{-q} > 0$. Thus, finding an \textit{inaccurate} query is identifying an index $i \in [2k]$ where $\ip{H - h}{s_i}$ is large, and higher $\ip{H - h}{s_i}$ means that $s_i$ is more informative for updating $H$. Inspired by this, the iterative selection process (\textsf{SelectMW}) proceeds in $T$ rounds:
\begin{enumerate}[leftmargin=*, label=\arabic*.]
  \item \textbf{Initialization:} Initialize the uniform distribution $p_1 = (1 / d, \cdots, 1 / d)\in \mathbb{R}^d$, so the candidate histogram is then $H_1 = N p_1$.
  \item \textbf{Selection via Exponential Mechanism:} In round $t \in [T]$, given the current model $H_t = N p_t$ and the actual histogram $h \in \R^d$, we sample a query index $i_t \in [2k]$ from the softmax distribution parameterized by $\lambda > 0$:
  \begin{equation}\label{eq:overview-softmax}
    P_h(i_t = i \mid i_1, \dots, i_{t-1}) = \frac{\exp\left( \lambda \ip{H_t - h}{s_i} \right)}{\sum_{a=1}^{2k} \exp\left( \lambda \ip{H_t - h}{s_a} \right)}.
  \end{equation}
  A larger $\lambda$ concentrates selection on queries where the current model $H_t$ deviates most from $h$.
  \item \textbf{Multiplicative Update:} Crucially, we \emph{do not} take a noisy measurement of $s_{i_t}(h)$ to determine the update step. Instead, we perform a fixed-step linear multiplicative update using only the selected vector $s_{i_t} \in [-1, 1]^d$:
  \begin{equation}\label{eq:overview-mw-update}
    \bar{p}_{t+1, j} = {p}_{t, j} \big( 1 - \rho s_{i_t}(j) \big), \qquad \forall j \in [d],
  \end{equation}
  where $\rho \in (0, 1/2]$ is a public learning rate. We then normalize $p_{t+1} = \bar{p}_{t+1} / \|\bar{p}_{t+1}\|_1$ and set $H_{t+1} = N p_{t+1}$. If $s_{i_t}(j) > 0$, the model over-predicted along coordinate $j$, so the factor $(1 - \rho s_{i_t}(j))$ decreases its weight; if $s_{i_t}(j) < 0$, its weight is increased.
\end{enumerate}

\paragraph{Deterministic Replay Decoding.}
A key consequence of update~\eqref{eq:overview-mw-update} is that every iterate $p_t$ is a deterministic function of the \emph{transcript} of queries selected by the exponential mechanism, $y = (i_1, \dots, i_T) \in [2k]^T$, and of the public parameters. So the update step is post-processing of private outputs and does not read $h$. Consequently, once the transcript $y$ is released, anyone can publicly replay the $T$ updates from $p_1 = \Unif([d])$ and compute the final synthetic distribution:
\begin{equation}\label{eq:overview-decoder}
  \widehat{p}  \coloneqq \frac{1}{T} \sum_{t=1}^T p_t \in \Delta_d.
\end{equation}
Therefore, releasing $\widehat{p}$ incurs no additional privacy loss beyond that of the transcript $y$. So, in the remainder of this section, we concentrate only on the privacy loss of outputting $y = (i_1, \dots, i_T)$ in the exponential mechanism selection~\eqref{eq:overview-softmax} over $T$ rounds.

\subsection{The Utility Trade-off and the Composition Barrier}
\label{sec:overview-barrier}

To evaluate the utility of $\widehat{p}$, we decompose the total error into two opposing components:
\begin{enumerate}[leftmargin=*]
  \item \textbf{MW Optimization Regret:} By tracking the standard relative entropy potential $\Phi_t = \KL(h/N \,\|\, p_t)$ starting from $\Phi_1\leq \log d$, a standard telescoping analysis ensures that for \emph{any} arbitrary sequence of signed queries, the realized empirical progress satisfies:
  \[
    \frac{1}{T} \sum_{t=1}^T \ip{p_t - \frac{h}{N}}{s_{i_t}} \le \frac{\log d}{\rho T} + \rho.
  \]
  Optimizing the learning rate by setting $\rho \asymp \sqrt{(\log d)/T}$ yields an error of $O\big(\sqrt{(\log d)/T}\big)$.
  \item \textbf{Softmax Selection Gap:} In each round $t$, the exponential mechanism samples $s_{i_t}$ instead of the exact worst-case query. By the standard analysis~\cite{MT07}, the sub-optimality gap between the maximum score and the expected selected score is bounded by:
  \[
    \max_{i \in [2k]} \ip{H_t - h}{s_{i}} - \E_{i_t}\left[ \ip{H_t - h}{s_{i_t}} \right] \le \frac{\log (2k)}{\lambda}.
  \]
  So this incurs a penalty of $\frac{\log(2k)}{N\lambda}$ per round after normalization.
\end{enumerate}
Combining these two components gives the ideal utility balance:
\begin{equation}\label{eq:overview-ideal-balance}
  \err_{\cQ}(\widehat p, h) \lesssim \sqrt{\frac{\log d}{T}} + \frac{\log(2k)}{N\lambda}.
\end{equation}

\paragraph{The Composition Barrier of Prior Work.}
Now consider differential privacy. In round $t$, changing one individual's record alters the histogram $h$ by $e_a - e_b$, which changes the inner product $\ip{H_t - h}{s_i}$ by at most $\|e_a - e_b\|_1 \|s_i\|_\infty \le 2$. The score function in \eqref{eq:overview-softmax} therefore has sensitivity $\Delta \le 2$. Again by standard analysis~\cite{MT07}, a single round satisfies pure $4\lambda$-DP. Therefore, the linear adaptive composition~\cite{DMNS06} dictates that the cumulative privacy loss over $T$ rounds is the linear sum: $\eps_{\mathrm{total}} =4\lambda T$. To guarantee pure $\eps$-DP overall, one is forced to set $\lambda \le \frac{\eps}{4T}$. Substituting this constraint into the utility balance~\eqref{eq:overview-ideal-balance} yields:
\begin{equation}\label{eq:overview-cubic-balance}
  \err_{\cQ} \lesssim \sqrt{\frac{\log d}{T}} + \frac{T \log(2k)}{N\eps}.
\end{equation}
The trade-off in \eqref{eq:overview-cubic-balance} is fatal: the second term grows linearly with $T$. Balancing the two forces requires choosing $T \asymp (N\eps)^{2/3}$, which locks the final error to the cube-root scaling $O\big( (N\eps)^{-1/3} \big)$. This linear composition penalty has been the universal barrier preventing iterative pure-DP algorithms (such as MWEM~\cite{HLM12}) from achieving the square-root rate.

\subsection{A Continuous Perspective: Pure DP via Logarithmic Derivatives}
\label{sec:overview-continuous}

To bypass step-by-step composition, we instead must analyze the privacy of the complete transcript $y \in [2k]^T$ as an \textit{integrated} object. Specifically, let $K = 2k$, for any fixed transcript $y = (i_1, \dots, i_T)$, let $p_v(y)$ denote the probability that \textsf{SelectMW} produces transcript $y$ when it is executed on an input vector $v \in \R^d$:
\begin{equation}\label{eq:overview-pv-def}
  p_v(y) = \prod_{t=1}^T P_v(i_t \mid i_1, \dots, i_{t-1}) = \prod_{t=1}^T \frac{\exp\left( \lambda \ip{H_t - v}{s_{i_t}} \right)}{\sum_{a=1}^K \exp\left( \lambda \ip{H_t - v}{s_a} \right)}.
\end{equation}
Although the true database $h$ is a discrete integer histogram in $\N^d$, the expression~\eqref{eq:overview-pv-def} is well-defined, strictly positive, and infinitely \textit{differentiable} for {all} real vectors $v \in \R^d$.

This continuous extension provides a powerful geometric lens. In \cref{def:dp-intro}, pure $\eps$-DP requires that for any neighboring histograms $h, h'$ differing by $\|h' - h\|_1 \le 2$, and for every transcript $y$,
\[
  \left| \log p_h(y) - \log p_{h'}(y) \right| \le \eps.
\]
By the fundamental theorem of calculus, if $h$ and $h'$ differ only in the $j$-th coordinate\footnote{Note that the replacement adjacency is equivalent to a group of two single-coordinate differences.}, the log-likelihood ratio can be expressed as the integral of the partial derivative along the line segment connecting them:
\begin{equation}\label{eq:overview-gradient-dp}
  \left|\log \frac{p_{h'}(y)}{p_h(y)}\right| 
  = \left| \int_0^1 \partial_{v_j} \log p_{h + \tau(h' - h)}(y) \cdot (h'_j - h_j) \, d\tau \right|
  \le \sup_{w \in \R^d} \left|\partial_{w_j} \log p_w(y)\right| \cdot |h'_j - h_j|.
\end{equation}

Thus, a uniform bound on the coordinate log-derivatives of the output probability is sufficient for pure differential privacy under replacement adjacency. However, a direct examination of $\nabla_v \log p_v(y)$ does not really help: it just confirms the composition barrier. In particular, let $P_t(i_t) = P_v(i_t \mid i_{<t})$, differentiating $\log P_t(i_t)$ with respect to coordinate $v_j$ again yields:
\[
  \partial_{v_j} \log P_t(i_t) = -\lambda s_{i_t}(j) + \lambda \E_{a \sim P_t}[s_a(j)].
\]
Because each query satisfies $s(j) \in [-1, 1]$, this single-round derivative is bounded by $2\lambda$. Summing over $T$ rounds gives:
\begin{equation}\label{eq:overview-first-deriv-bad}
  \left| \partial_{v_j} \log p_v(y) \right| \le 2\lambda T.
\end{equation}
In the worst-case transcript where the selected queries $s_{i_t}$ consistently align in the same direction across all $T$ rounds, the slope $|\partial_{v_j} \log p_v(y)|$ genuinely scales as $\Theta(\lambda T)$, so we go back to $\lambda = O(\eps/T)$ from basic composition.


The starting point of our approach begins with a simple observation: while the \emph{first} derivative of $\log p_v(y)$ can accumulate linearly to $\Theta(T\lambda)$, its \emph{second} derivative exhibits a dramatic structural rigidity. Fix a transcript $y$ and differentiate $\log P_t(i_t)$ a second time with respect to $v_j$. 
A direct calculation yields:
\begin{align}
  \partial_{v_j}^2 \log P_t(i_t) 
  &= \partial_{v_j} \left( -\lambda s_{i_t}(j) + \lambda \frac{\sum_{a=1}^K s_a(j) e^{\lambda \ip{H_t - v}{s_a}}}{\sum_{a=1}^K e^{\lambda \ip{H_t - v}{s_a}}} \right) \notag\\
  &= -\lambda^2 \left( \frac{\sum_a s_a(j)^2 e^{\lambda \ip{H_t-v}{s_a}}}{\sum_a e^{\lambda \ip{H_t-v}{s_a}}} - \left(\frac{\sum_a s_a(j) e^{\lambda \ip{H_t-v}{s_a}}}{\sum_a e^{\lambda \ip{H_t-v}{s_a}}}\right)^2 \right) \notag\\
  &= -\lambda^2 \Var_{a \sim P_t}[s_a(j)]. \label{eq:overview-variance}
\end{align}
We could see that the second derivative of the log-probability is exactly the negative variance of the query selection under the distribution defined by exponential mechanism. Since $s_a(j) \in [-1, 1]$, the variance of any bounded random variable supported on $[-1, 1]$ is trivially at most $1$. Therefore, for every round $t$, we have the one-sided curvature bound:
\[
  \partial_{v_j}^2 \log P_t(i_t) = -\lambda^2 \Var_{a \sim P_t}[s_a(j)] \ge -\lambda^2.
\]
Summing this relation across all $T$ rounds of the transcript yields a global, uniform lower bound on the Hessian:
\begin{equation}\label{eq:overview-curvature-bound}
  \partial_{v_j}^2 \log p_v(y) \ge -T \lambda^2, \qquad \forall v \in \R^d, \quad \forall j \in [d], \quad \forall y \in [K]^T.
\end{equation}
Note that this bound holds \emph{pointwise} across the entire space $\R^d$. Geometrically, it asserts that the function $v \mapsto \log p_v(y)$ is coordinate-wise \textbf{semi-convex}: along every coordinate axis, its downward curvature is strictly bounded by $r^2 \coloneqq T\lambda^2$. It is physically impossible for the log-probability function to curve downward more sharply than a fixed parabola of curvature $-T\lambda^2$.

\subsection{Intuition: Steep Slopes Imply Distant Mountains}
\label{sec:overview-mountain-intuition}

How can a lower bound on the \emph{second} derivative in \eqref{eq:overview-curvature-bound} help us control the \emph{first} derivative and guarantee pure differential privacy? We illustrate this by a scalar function as an example. Let $f(x) \coloneqq p_v(y)$ denote the probability of a fixed transcript $y$ along a one-dimensional coordinate line $x = v_j$, and let $\psi(x) \coloneqq \log f(x)$. Our curvature bound guarantees that $\psi''(x) \ge -r^2$ everywhere, where $r = \lambda \sqrt{T}$.

Suppose at some point $x$, the log function exhibits an extremely steep slope, say $|\psi'(x)| = s \gg r$, which violates pure DP. Assume without loss of generality that $\psi'(x) = +s > 0$.
Recall that the second derivative represents the \emph{rate of change of the slope}. The condition $\psi''(x) \ge -r^2$ imposes a strict speed limit on how fast the slope can decrease:
\[
  \psi'(x + t) = \psi'(x) + \int_0^t \psi''(x + \tau) \, d\tau \ge s - r^2 t.
\]
Notice what this means: the slope \emph{cannot instantly drop back to zero}. As we move to the right ($t > 0$), the slope remains strictly positive for all $t < s/r^2$. Integrating once more from $0$ to $t$ gives a quadratic lower bound on the log-probability:
\begin{equation}\label{eq:overview-quadratic-surge}
  \psi(x + t) \ge \psi(x) + s t - \frac{1}{2} r^2 t^2.
\end{equation}
Exponentiating back to the original probability $f(x) = e^{\psi(x)}$ reveals:
\begin{equation}\label{eq:overview-f-mountain}
  f(x + t) \ge f(x) \exp\left( s t - \frac{1}{2} r^2 t^2 \right).
\end{equation}
The quadratic in the exponent attains its maximum at $t^* = s/r^2$. At this peak, the exponent reaches $s^2 / (2r^2)$. 
Therefore, at a distance of $t^* \approx s/r^2$ to the right, the function value satisfies:
\[
  f(x + t^*) \ge f(x) \cdot \exp\left( \frac{s^2}{2r^2} \right).
\]
Thus, intuitively, a steep slope $s$ cannot be a transient, isolated spike. The curvature lower bound $\psi'' \ge -r^2$ forces the function to undergo a sustained, quadratic surge, culminating in a point whose height is magnified by the super-exponential factor $\exp(s^2 / (2r^2))$ relative to $f(x)$.

As a consequence, the geometric insight is that, if we perform a random walk (i.e., diffusion) starting from $x$, the probability density will easily sample points along this massive mountain, discovering function values that dominate the base value $f(x)$.

\paragraph{Laplace Convolution and Its Inherent Smoothness.}
This geometric insight inspires a \textit{regularizing} procedure. If $f(x)$ has large slopes, so that $|(\log f(x))'|$ may be large so the pure-DP condition is violated, we can regularize it by forming a mixture with a diffused version of itself. Specifically, define $g(x)$ to be the convolution of $f$ with a centered Laplace density $l(t) = \frac{\kappa}{2} e^{-\kappa |t|}$:
\begin{equation}\label{eq:overview-g-def}
  g(x) \coloneqq (l * f)(x) = \int_{-\infty}^\infty \frac{\kappa}{2} e^{-\kappa |t|} f(x + t) \, dt.
\end{equation}
The convolution $g(x)$ represents the expected value of $f(x + Z)$ where $Z \sim \Lap(\kappa)$ is a random walk step drawn from a Laplace distribution.

The Laplace kernel possesses a property that its derivative satisfies $|l'(t)| \le \kappa l(t)$ almost everywhere. Using this fact, we show that for the convolution:
\begin{equation}\label{eq:overview-g-smooth}
  |g'(x)| \le \kappa g(x) \quad \Longleftrightarrow \quad \left| \frac{d}{dx} \log g(x) \right| \le \kappa.
\end{equation}
That is, the function $g(x)$ is \emph{universally Lipschitz in log-space}, with logarithmic slope naturally bounded by $\kappa$, independent of $f$.

\paragraph{The Composite Mixture.}
Now consider a mixture that retains the original function $f(x)$ with large probability $1 - \pi$, and invokes the Laplace random walk $g(x)$ with a tiny probability $\pi \in (0, 1)$:
\begin{equation}\label{eq:overview-H-def}
  H(x) \coloneqq (1 - \pi) f(x) + \pi g(x).
\end{equation}
We then show how does this mixture behave at a point $x$ where $f$ has steep logarithmic slope 
$$s = |\psi'(x)| = \left|\frac{d\log f(x)}{dx}\right| =|f'(x)|/f(x).$$
Note that the derivative of the mixture is $H'(x) = (1 - \pi) f'(x) + \pi g'(x)$. To measure the privacy loss, consider its logarithmic slope:
\begin{equation}\label{eq:overview-H-ratio}
  \left| \frac{d \log H(x)}{dx}\right| = \frac{|H'(x)|}{H(x)} = \frac{|(1 - \pi) f'(x) + \pi g'(x)|}{(1 - \pi) f(x) + \pi g(x)} \le \frac{(1 - \pi) s f(x) + \pi \kappa g(x)}{(1 - \pi) f(x) + \pi g(x)}.
\end{equation}
This is a weighted average between the slope $s$ and the safe slope $\kappa$. 
In \cref{prop:scalar-smoothing}, we will show that, for moderate slope, that is, if $s \le \kappa + r u$ (where $r^2 = T\lambda^2$ and $u \approx \sqrt{2\log(1/\pi)}$), then both components are already well-behaved, and the ratio~\eqref{eq:overview-H-ratio} is trivially bounded by $\kappa + ru$.

For the case of steep slope where $s > \kappa + ru$, by our geometric insight from Section~\ref{sec:overview-mountain-intuition}, this steep logarithmic derivative ensures the presence of a pronounced peak of magnitude at least $f(x) \exp(s^2 / (2r^2))$ within a short distance. When the Laplace kernel samples the domain, its integral over this peak blows up:
  \[
    \frac{g(x)}{f(x)} \ge \frac{1}{4} \exp\left( \frac{(s - \kappa)^2}{2r^2} \right).
  \]
  Because $s$ is large, this exponential surge easily overwhelms the mixture coefficient $\pi$, in which case $\pi g(x)$ completely dominates $(1 - \pi) f(x)$:
  \[
    \pi g(x) \gg (1 - \pi) f(x).
  \]
  Thus, the convolution term $g(x)$ takes over the denominator of \eqref{eq:overview-H-ratio}. The term $(1 - \pi) s f(x)$ in the numerator is absorbed and diluted by the larger mass of $\pi g(x)$, pulling the overall mixed ratio $|H'(x)/H(x)|$ back down to the target bound $\kappa + ru$. 

Motivated by the above intuition, we establish the following. For any twice differentiable function $f \colon \mathbb{R} \to (0, 1]$ with $(\log f)''(x) \ge -r^2$, any $\pi \in (0, 1)$, any $\kappa \ge r$, and $l(t) = \frac{\kappa}{2}e^{-\kappa|t|}$, recall that $H(x)$ is the mixed function defined by $$H(x) = (1 - \pi)f(x) + \pi g(x) = (1 - \pi)f(x) + \pi (l * f)(x).$$ Setting $u = \sqrt{2\log(4/\pi)}$ and $L = \kappa + ru$, we then have, for all $x \in \mathbb{R}$,
  \[
  \left| \frac{d \log H(x)}{dx}\right| = |H'(x)/ H(x)| \leq L,
\]
which implies that for all $t\in \mathbb{R}$,
\begin{equation}\label{eq:intro-1d}
      H(x + t) \le e^{L|t|} H(x).
\end{equation}
The formal proof is given in Appendix~\ref{app:scalar-smoothing}.

\subsection{The Preprocessing: Sparse Laplace Smoothing}
\label{sec:overview-preprocessing}
To summarize the discussion thus far, consider a function $f(x) \in (0,1]$ representing a one-dimensional slice of the transcript distribution $p_v(y)$ along a single input coordinate $x \in \R$. By integration, the logarithmic derivative $\abs{(\log f)'(x)}$ naturally dominates the pure differential privacy loss $\abs{\log \frac{p_v(y)}{p_{v'}(y)}}$ between neighboring datasets $v \sim v'$. Although this unperturbed slope $\abs{(\log f)'(x)}$ can be large in the worst case, a uniform lower bound on the log-curvature, $(\log f)''(x) \ge -r^2$, guarantees that the slope cannot drop too rapidly. Consequently, taking a convex mixture of $f(x)$ with its Laplace convolution $g(x) = \int_{-\infty}^\infty \frac{\kappa}{2} e^{-\kappa \abs{t}} f(x + t) \, dt$ produces a regularized function $H(x) \coloneqq (1-\pi)f(x) + \pi g(x)$ that exhibits a globally controlled logarithmic slope $\abs{(\log H)'(x)} \le \kappa + O(r)$. 

From an algorithmic perspective, this functional mixture translates directly into a pre-processing step: with probability $\pi$, we perturb the coordinate $x$ by adding independent Laplace noise $Z \sim \Lap(\kappa)$ and evaluate the mechanism on $x + Z$; otherwise, the input coordinate is left untouched.

With this one-dimensional analytic principle established, our full algorithm is rather simple. We do not alter the inner workings of \textsf{SelectMW}, nor do we post-process outputs. Instead, we apply a single randomized preprocessing step directly to the input histogram $h \in \N^d$:

\begin{center}
  \begin{tcolorbox}[sharpish corners, colback=white, width=1\linewidth]
    \refstepcounter{procedure}\label{proc:overview-alg}
    \begin{center}
      \textbf{Selection-only MW with Input Smoothing}
    \end{center}
    \vspace{6pt}
Independently for each coordinate $j \in [d]$:
\begin{itemize}[leftmargin=*]
  \item Draw a Bernoulli indicator $B_j \sim \operatorname{Bernoulli}(\pi)$ with probability $\pi = \frac{\beta}{32d}$.
  \item If $B_j = 1$, draw independent Laplace noise $Z_j \sim \Lap(\kappa)$ with rate $\kappa = \eps / 8$.
  \item If $B_j = 0$, set $Z_j = 0$.
\end{itemize}
Let $\widetilde{h} \coloneqq h + Z \in \R^d$ be the perturbed vector, and execute \textsf{SelectMW}($\widetilde{h}$) across all $T$ rounds.
  \end{tcolorbox}
\end{center}
The induced distribution over transcripts under this smoothed input is precisely the mixture:
\[
  Q_h(y) = \E_Z\big[ p_{h + Z}(y) \big].
\]
To verify pure $\eps$-DP, consider replacing a single record in $h$, leading to an adjacent histogram $h'$ with $\|h' - h\|_1 \le 2$. We interpolate from $h$ to $h'$ coordinate-by-coordinate. Fixing any transcript $y$, an intermediate vector $v$ along the interpolation, and conditioning on the perturbations $z_{-j}$ on all other coordinates, the 1D slice $f(x) \coloneqq p_{(v_{-j}+z_{-j}, x)}(y)$ satisfies $(\log f)''(x) = \partial_{w_j}^2 \log p_w(y)$ at $w = (v_{-j} + z_{-j}, x)$. According to \eqref{eq:overview-curvature-bound}, this partial second derivative is uniformly lower-bounded across $\R^d$ by 
$$-\lambda^2 \sum_{t=1}^T \Var_{i \sim P_t}[s_i(j)] \ge -T\lambda^2 := -r^2,$$ so the slice $f$ identically satisfies the 1D curvature lower bound with $r = \lambda\sqrt{T}$. Then, applying \eqref{eq:intro-1d} to each coordinate shift and integrating over the remaining latent noise establishes that for any event $F \subseteq [K]^T$:
\begin{equation}\label{eq:overview-multivariate-privacy}
  Q_{h'}(F) \le \exp\left( L \|h' - h\|_1 \right) Q_h(F) \le e^{2L} Q_h(F),
\end{equation}
where $L = \kappa + r\sqrt{2\log(4/\pi)}$. 
By choosing $\kappa = \eps / 8$, $\pi = \frac{\beta}{32 d}$ and calibrating $\lambda$ such that:
\begin{equation}\label{eq:overview-lambda-setting}
  r \sqrt{2\log(4/\pi)} = \lambda \sqrt{T} \sqrt{2\log(4/\pi)} \le \frac{\eps}{8} \quad \Longleftrightarrow \quad \lambda \asymp \frac{\eps}{\sqrt{T \log(d/\beta)}},
\end{equation}
we obtain $2L \le 2(\eps/8 + \eps/8) = \eps/2 \le \eps$. 
Thus, {the smoothed transcript distribution $Q_h$ satisfies pure $\eps$-differential privacy} while we can choose $\lambda \sim T^{{-1/2}}$ instead of $\lambda \sim T^{-1}$ from basic composition. 

Regarding utility, it is easy to verify that the total variation distance satisfies $\norm{Q_h - p_h}_{\TV} \le d\pi \le \beta/32$. Consequently, any high-probability utility guarantee established for the original process $p_h$ carries over to the private mechanism $Q_h$ with at most an additive $O(\beta)$ penalty in failure probability. In particular, on the clean event where no coordinate is perturbed ($Z = 0$), as we have already discussed in \cref{sec:overview-barrier}, the standard multiplicative-weights potential analysis and martingale concentration guarantee an optimization error of $O(\sqrt{(\log d)/T})$. Combining this with the exponential mechanism's selection gap, the overall query error satisfies:
\begin{equation*}
  \err_{\cQ}(\widehat p, h) \lesssim \sqrt{\frac{\log d}{T}} + \frac{\log(2k)}{N\lambda} \lesssim \sqrt{\frac{\log(d/\beta)}{T}} + \frac{\sqrt{T \log(d/\beta)} \log(2k)}{N\eps}
\end{equation*}
with high probability. Crucially, the second term now grows with $\sqrt{T}$ rather than the linear $T$ incurred by step-by-step composition. Balancing these two opposing forces by setting the number of rounds to $T \asymp \lceil N\eps / \log(2k) \rceil$ yields the square-root rate
\[
  \err_{\cQ}(\widehat p, h) \le O\!\left( \sqrt{\frac{\log(2d/\beta)\log(2k)}{N\eps}} \right),
\]
matching the statement of \cref{thm:hp}.
\begin{remark}
    We note that the Laplace noise $Z$ is \textbf{not} added to privatize the input histogram $h$. Here, $Z$ is a purely smoothing operator acting on the log-likelihood function $v \mapsto \log p_v(y)$. Its sole purpose is to regularize the derivative of the \emph{induced output distribution} $Q_h(y)$.
\end{remark}

\paragraph{A Conceptual Takeaway: From Stability to the Stability of Stability.}
At a conceptual level, standard differential privacy analyses are usually first-order: they characterize the algorithmic \emph{stability} of an output distribution under perturbations of the input database, which translates to bounding the first derivative of the log-likelihood function $\abs{\partial_{v_j} \log p_v(y)}$. Traditional pure-DP iterative mechanisms hit a bottleneck because worst-case adaptive composition~\cite{DMNS06} forces these first derivatives to accumulate linearly at rate $\Theta(T)$.

In contrast, our framework pivots to analyzing the \emph{stability of stability}: the second-order curvature $\partial_{v_j}^2 \log p_v(y)$ that dictates how the local sensitivity itself fluctuates across the data domain. By establishing that this log-curvature is lower-bounded by negative variances across iterations ($\ge -T\lambda^2$), we uncover a geometric intuition: the first-order privacy loss cannot surge arbitrarily without creating a sustained, macroscopic surge in probability mass in an immediate neighborhood. This higher-order perspective reveals that destructive per-round noise addition is unnecessary. Instead, by understanding the fine-grained evolution of stability itself, a single, sparse input smoothing suffices to globally tame the first-order privacy loss, finally achieving the optimal square-root error rate in polynomial time.


\section{Preliminaries}\label{sec:preliminaries}

Write \(\N=\{0,1,2,\ldots\}\), probability simplex \(\Delta_d=\{p\in\R_{\geq0}^d:\sum_jp_j=1\}\), and feasible set of histograms \(\mathcal H_{N,d}=\{h\in\N^d:\sum_jh_j=N\}\). All logarithms are natural unless their base is displayed.
The query table, \(N,d,k\), and the privacy and confidence parameters
are public.

\subsection{Basics of Differential Privacy}\label{sec:dp-basics}

\begin{definition}[Replacement adjacency and differential privacy]
\label{def:dp}
Two histograms \(h,h'\in\mathcal H_{N,d}\) are adjacent, written
\(h\sim h'\), if \(h'=h+e_a-e_b\) for some \(a,b\in[d]\), where
\(e_j\) is the \(j\)-th standard basis vector.
For \(\eps\geq0\) and \(0\leq\delta<1\), a randomized mechanism
\(\mathcal M\) is \((\eps,\delta)\)-differentially private if, for
every adjacent pair and every measurable output event \(F\),
\[
 \Prb\{\mathcal M(h)\in F\}
 \leq e^\eps\Prb\{\mathcal M(h')\in F\}+\delta.
\]
We call \((\eps,0)\)-DP \emph{pure} \(\eps\)-DP.
\end{definition}
Adjacency is symmetric, so interchanging the inputs gives the reverse
comparison.  In particular, \(\norm{h-h'}_1\leq2\) for adjacent
histograms.  Our query-release mechanism satisfies pure privacy;
\(\beta\) always denotes an accuracy failure probability.

\begin{lemma}[Post-processing~\cite{DR14}]
\label{lem:postprocess}
If \(\mathcal M\) is \((\eps,\delta)\)-DP and \(g\) uses only its
output, public information, and fresh randomness, then
\(g\circ\mathcal M\) is \((\eps,\delta)\)-DP.
\end{lemma}

\begin{lemma}[Adaptive composition for pure privacy~\cite{DR14}]
\label{lem:composition}
At each step \(t\in[T]\), a mechanism \(\mathcal M_t\) may depend
on the previous outputs \(y_{<t}\).  Suppose that, for every fixed
history, \(\mathcal M_t(\cdot;y_{<t})\) is pure \(\eps_t\)-DP,
where the budgets \(\eps_t\) are fixed in advance.
Then the joint transcript is pure \((\sum_{t=1}^T\eps_t)\)-DP.
\end{lemma}

\begin{definition}[Sensitivity]\label{def:sensitivity}
The \(\ell_1\)-sensitivity of a vector query
\(f:\mathcal H_{N,d}\to\R^m\) is
\[
 \Delta_1(f)=\sup_{h\sim h'}\norm{f(h)-f(h')}_1.
\]
For a scoring function \(u:\mathcal H_{N,d}\times\mathcal Y\to\R\), its
sensitivity is \(\Delta_u=\sup_{h\sim h',\,y\in\mathcal Y}
|u(h,y)-u(h',y)|\).
\end{definition}

We use \(\Lap(\kappa)\) for the centered Laplace distribution with
\emph{rate} \(\kappa>0\), whose density is
\((\kappa/2)e^{-\kappa|z|}\).  Thus its scale is \(1/\kappa\).

\begin{lemma}[Laplace mechanism~\cite{DMNS06}]
\label{lem:laplace-priv}
Let \(f:\mathcal H_{N,d}\to\R^m\) satisfy
\(\Delta_1(f)\leq\Delta\), where \(\Delta>0\).
For \(\eps>0\), releasing \(f(h)+(Z_1,\ldots,Z_m)\), with
independent \(Z_j\sim\Lap(\eps/\Delta)\), is pure \(\eps\)-DP.
\end{lemma}
A query of sensitivity zero can be released exactly.  In particular,
one normalized query \(q\in[0,1]^d\) has sensitivity at most \(1/N\),
and the vector of all \(k\) answers has \(\ell_1\)-sensitivity at
most \(k/N\).

\begin{lemma}[Exponential mechanism~\cite{MT07}]
\label{lem:exponential-mechanism}
Let \(\mathcal Y\) be a finite nonempty public set, and let
\(u(h,y)\) be a score of sensitivity at most \(\Delta\).
For \(\lambda>0\), sample \(Y\) with probabilities
\[
 \Prb\{Y=y\}
 =\frac{e^{\lambda u(h,y)}}{\sum_{z\in\mathcal Y}e^{\lambda u(h,z)}}.
\]
This mechanism is pure \(2\lambda\Delta\)-DP.  For every fixed
input and \(0<\theta<1\), with probability at least \(1-\theta\),
\[
 u(h,Y)\geq\max_{y\in\mathcal Y}u(h,y)
                 -\frac{\log(|\mathcal Y|/\theta)}{\lambda}.
\]
\end{lemma}

\subsection{Probability Tools}\label{sec:basic-tools}

\begin{lemma}[Laplace tails]\label{lem:laplace-tail}
If \(Z\sim\Lap(\kappa)\), then for every \(t\geq0\),
\[
 \Prb\{|Z|>t\}=e^{-\kappa t}.
\]
Consequently, for \(m\) variables with these marginals and
\(0<\theta<1\),
\[
 \Prb\!\left\{\max_{j\in[m]}|Z_j|
       >\frac{\log(m/\theta)}{\kappa}\right\}\leq\theta.
\]
This bound does not require independence.
\end{lemma}
\begin{proof}
Integrating the density outside \([-t,t]\) gives the first identity;
a union bound gives the maximum bound.
\end{proof}

For laws \(P,Q\) on a finite set \(\mathcal Y\), write
\(\norm{P-Q}_{\TV}=\frac12\sum_{y\in\mathcal Y}|P(y)-Q(y)|\).
We use the following facts to compare the original and smoothed
transcript laws.

\begin{lemma}[Total variation bounds~\cite{jerrum1998mathematical}]\label{lem:tv}
For laws \(P,Q\) on a finite set,
\[
 \norm{P-Q}_{\TV}=\max_{F\subseteq\mathcal Y}|P(F)-Q(F)|.
\]
In particular, total variation bounds the error at each atom and
each event, and satisfies the triangle inequality.
\end{lemma}

\begin{lemma}[Coupling~\cite{jerrum1998mathematical}]\label{lem:coupling}
For any coupling \((X,Y)\) of finite laws \(P,Q\),
\(\norm{P-Q}_{\TV}\leq\Prb\{X\ne Y\}\), and a coupling attaining
equality exists.
\end{lemma}

For probability vectors \(p,q\) with all \(q_j>0\), define
\(\KL(p\|q)=\sum_{j:p_j>0}p_j\log(p_j/q_j)\).

\begin{lemma}[Relative entropy bounds~\cite{vershynin2019high}]\label{lem:kl}
For probability vectors \(p,q\in\Delta_m\) with \(q>0\),
\(\KL(p\|q)\geq0\).  In addition,
\[
 0\leq-\sum_{j:p_j>0}p_j\log p_j\leq\log m,
 \qquad
 \KL(p\|\Unif([m]))\leq\log m.
\]
\end{lemma}

\section{Input Laplace Smoothing for Iterative Mechanisms}\label{sec:smoothing}
The generic smoothing lemma (\cref{lem:multivariate-smoothing}) established in this section serves as our central analytic ingredient. In this section, we consider an iterative algorithm with $T$ rounds in a generic setting. In particular, let $v\in \mathbb{R}^d$ be the input\footnote{We remark that in our actual setting, the coordinates of the histograms are integers, so the setting stated here is in fact more general.} and $y\in \mathcal{Y}^T$ be the output sequence. We consider the following iterative procedure.

\paragraph{Original iterative procedure.}
Let $\mathcal{M}_t:\R^d\times\mathcal{Y}^{t-1}\to\mathcal{Y}$ for $t\in[T]$ be a series of randomized steps. Each step uses fresh randomness and may depend on the preceding outputs $y_{<t}=(y_1,\ldots,y_{t-1})$. 
\begin{center}
  \begin{tcolorbox}[sharpish corners, colback=white, width=1\linewidth]
    \refstepcounter{procedure}\label{proc:original-procedure}
    \begin{center}
      \textbf{\emph{Procedure \ref*{proc:original-procedure}: Original iterative procedure $\mathcal{M}(v)$}}
    \end{center}
    \vspace{6pt}
    \begin{enumerate}
      \item \textbf{Input:} $v\in\R^d$; public transition rules $\mathcal{M}_1,\ldots,\mathcal{M}_T$.
      \item \textbf{For} each $t=1,\ldots,T$:
      \begin{itemize}
        \item Draw $y_t\leftarrow\mathcal{M}_t(v,y_{<t})$.
      \end{itemize}
      \item \textbf{Output:} The transcript $y=(y_1,\ldots,y_T)$.
    \end{enumerate}
  \end{tcolorbox}
\end{center}
The final distribution is defined by
\begin{equation}\label{eq:original-transcript-law}
    p_v(y)\coloneqq\Prb[\mathcal{M}(v)=y]
    =\prod_{t=1}^T\Prb[\mathcal{M}_t(v,y_{<t})=y_t],
    \qquad y\in\mathcal{Y}^T.
\end{equation}
Throughout this section, for every fixed transcript $y$, we assume that
$v\mapsto p_v(y)$ is strictly positive and twice continuously
differentiable on $\R^d$.

In the analysis of $T$-round differentially private iterative mechanisms with input $v$ and transcript $y$,  one frequently encounters these probability mass functions $p_v(y)$, and its log-likelihood ratio $\log \frac{p_v(y)}{p_{v'}(y)}$ between neighboring inputs $v \sim v'$ typically scales linearly with $T$ for pure differential privacy analysis, which is in general unavoidable under worst-case composition. If we view through a \emph{continuous relaxation} for $v \in \mathbb{R}^d$, this discrete privacy loss reflects the fact that the coordinate-wise derivative $\vert{}\partial_{v_j} \log p_v(y)\vert{}$ can accumulate at rate $\Theta(T)$ when $v$ and $v'$ differ along coordinate $j \in [d]$.

The purpose of this section is to show that this linear barrier can be bypassed under a mild uniform curvature assumption. Specifically, assuming that the Hessian of $\log p_v(y)$ admits a negative semidefinite lower bound $\nabla^2_v \log p_v(y) \succeq -r^2 I$ (or, more generally, coordinate-wise, $\partial_{v_j}^2 \log p_v(y) \geq -r^2$) for some $r > 0$, we construct a smoothed probability law $Q_v$ that is close to $p_v$ in total variation distance while guaranteeing that $\Vert{}\nabla \log Q_v(y)\Vert{}_\infty$ is controlled directly by $r$ instead of $T$.

\paragraph{Input-smoothed procedure.}
Fix $\pi\in(0,1)$ and $\kappa>0$. In each coordinate $j\in [d]$, with probability $\pi$, the smoothed procedure draws an independent Laplace perturbation\footnote{To prevent misunderstanding, as we will show later, the Laplace perturbation here is \emph{not} intended to privatize the input; rather, it is used to smooth the function $p_v(y)$.} with rate $\kappa$ before the first round and then simply invokes the original procedure on the
same perturbed input throughout all $T$ rounds.

  \begin{tcolorbox}[float=tp, sharpish corners, colback=white, width=1\linewidth]
    \refstepcounter{procedure}\label{proc:smoothed-procedure}
    \begin{center}
      \textbf{\emph{Procedure \ref*{proc:smoothed-procedure}: Input-smoothed procedure $\widetilde{\mathcal{M}}(v;\pi,\kappa)$}}
    \end{center}
    \vspace{6pt}
    \begin{enumerate}
      \item \textbf{Input:} $v\in\R^d$; original procedure $\mathcal{M}$; public $\pi\in(0,1)$ and $\kappa>0$.
      \item \textbf{For} each $j=1,\ldots,d$:
      \begin{itemize}
        \item Independently draw $B_j\sim\operatorname{Bernoulli}(\pi)$.
        \item If $B_j=1$, independently draw $Z_j\sim\Lap(\kappa)$; otherwise set $Z_j=0$.
      \end{itemize}
      \item Run original procedure $y\leftarrow\mathcal{M}(v+Z)$.
      \item \textbf{Output:} The transcript $y=(y_1,\ldots,y_T)$.
    \end{enumerate}
  \end{tcolorbox}
In particular, in \cref{proc:smoothed-procedure}, the input perturbation has independent coordinates with law
\begin{equation}\label{eq:sparse-noise-def}
    Z_j\sim(1-\pi)\delta_0+\pi\Lap(\kappa),\qquad j\in[d],
\end{equation}
where $\delta_0$ is the point mass at zero, and the smoothed transcript law is
\begin{equation}\label{eq:Q-def}
\begin{aligned}
    Q_v(y) &\coloneqq \Prb[\widetilde{\mathcal{M}}(v;\pi,\kappa)=y] \\
    &= \E_Z\big[\Prb[\widetilde{\mathcal{M}}(v;\pi,\kappa)=y \mid Z]\big] = \E_Z\big[\Prb[\mathcal{M}(v+Z)=y \mid Z]\big]\\
    & = \E_Z[p_{v+Z}(y)]
\end{aligned}
\end{equation}
for every $y\in\mathcal{Y}^T$. Since $p_v(y) > 0$ on all of $\R^d$, the random variable $p_{v+Z}(y)$ is strictly positive almost surely, ensuring $Q_v(y) = \E[p_{v+Z}(y)] > 0$. Further, interchanging the finite sum over $\mathcal{Y}^T$ with the expectation gives $\sum_{y \in \mathcal{Y}^T} Q_v(y) = \E\big[\sum_{y \in \mathcal{Y}^T} p_{v+Z}(y)\big] = \E[1] = 1$, which verifies that $Q_v$ is a well-defined probability law on $\mathcal{Y}^T$.

\begin{lemma}[Sparse Smoothing of an Iterative Procedure]
\label{lem:multivariate-smoothing}
Suppose the original transcript law $p_v(y)$ satisfies, for some
$0<r\leq\kappa$,
\begin{equation}\label{eq:lem-curvature-assumption}
    \partial_{v_j}^2\log p_v(y)\geq-r^2
    \quad\text{for all }v\in\R^d,\ j\in[d],\ y\in\mathcal{Y}^T.
\end{equation}
Set
\begin{equation}\label{eq:L-def}
    L\coloneqq\kappa+r\sqrt{2\log\left(\frac{4}{\pi}\right)}.
\end{equation}
Then, distribution $Q_v$ satisfies the following properties:
\begin{enumerate}[label=(\roman*)]
    \item \textbf{Pure privacy / Log-Lipschitz bound:} For all $v, v' \in \R^d$ and every event $F \subseteq \mathcal{Y}^T$,
    \begin{equation}\label{eq:event-privacy}
        Q_{v'}(F) \leq e^{L \norm{v' - v}_1} Q_v(F).
    \end{equation}
    \item \textbf{Pointwise domination:} For every $y \in \mathcal{Y}^T$,
    \begin{equation}\label{eq:pointwise-domination}
        Q_v(y) \geq (1-\pi)^d p_v(y).
    \end{equation}
    \item \textbf{Total variation proximity:} The total variation distance satisfies
    \begin{equation}\label{eq:tv-bound}
        \norm{Q_v - p_v}_{\TV} \leq d \pi.
    \end{equation}
\end{enumerate}
\end{lemma}

To prove \cref{lem:multivariate-smoothing}, we first isolate the core mechanism into a pure functional analysis proposition (\cref{prop:scalar-smoothing}), and subsequently derive the multivariate lemma as an immediate application. In particular, we analyze the regularizing effect of a sparse Laplace mixture on an arbitrary smooth function whose logarithmic curvature is bounded.

\begin{restatable}[Logarithmic Lipschitz Bound via Sparse Laplace Mixture]{proposition}{ScalarSmoothing}\label{prop:scalar-smoothing}
Let $f \colon \R \to (0, 1]$ be twice continuously differentiable with $(\log f)''(x) \geq -r^2$ for some $r > 0$.
For $\pi \in (0, 1)$, $\kappa \geq r$, and Laplace density $l(t) \coloneqq \frac{\kappa}{2} e^{-\kappa \abs{t}}$, define the mixture $$H(x) \coloneqq (1-\pi)f(x) + \pi \int_{-\infty}^\infty l(t) f(x+t) \, dt$$ and parameters
\begin{equation}\label{eq:prop-constants}
    u \coloneqq \sqrt{2 \log(4/\pi)} > 1, \qquad L \coloneqq \kappa + r u = \kappa + r \sqrt{2 \log(4/\pi)}.
\end{equation}
Then, $H(x)$ is continuously differentiable, and for every $x \in \R$ and $t \in \R$,
\begin{equation}\label{eq:prop-shift-bound}
    H(x+t) \leq e^{L \abs{t}} H(x).
\end{equation}
\end{restatable}
We have illustrated the proof sketch of \cref{prop:scalar-smoothing} in \cref{sec:overview-mountain-intuition}. The full proof of \cref{prop:scalar-smoothing} is deferred to Appendix~\ref{app:scalar-smoothing}. Intuitively, the convolution $$g(x) = l * f = \int_{-\infty}^\infty l(t) f(x+t) dt$$ with the Laplace kernel serves as a self-regularizing component, since $g$ itself has $|g'(x)| \leq \kappa g(x)$, ensuring that the convolution alone has a controlled logarithmic derivative. 

When the unperturbed slope $|(\log f)'(x)|$ is already moderate (no larger than $\kappa + ru$), both components of the mixture $H = (1-\pi)f + \pi g$ are well-behaved, and the target bound follows easily. 
When $|(\log f)'(x)|$ is larger than $\kappa + ru$, the uniform log-curvature lower bound $(\log f)'' \geq -r^2$ comes into play: it prevents the logarithmic slope from dropping too quickly, which forces $f$ to undergo a sustained quadratic surge of the form $f(x \pm t) \geq f(x)\exp(|(\log f)'(x)| t - r^2 t^2 / 2)$ along the ray of steep ascent for all $t \geq 0$. 
Integrating this quadratic growth against the Laplace kernel amplifies the convolution $g(x)$ by a factor that easily dominates $f(x)$, so that the tiny $\pi$-fraction of $g(x)$ absorbs the steep slope of $(1-\pi)f(x)$.  Consequently, the ratio $|H'(x)/H(x)| \leq L$ holds universally on $\R$, and integrating this logarithmic derivative yields the multiplicative guarantee $H(x+t) \leq e^{L|t|}H(x)$.

We now apply \cref{prop:scalar-smoothing} to analyze the smoothed probability law $Q_v(y)$ on the finite transcript space $\mathcal{Y}^T$. This multivariate result follows by conditioning on all coordinates except one, and then invoking \cref{prop:scalar-smoothing} on the resulting one-dimensional slice.

\begin{proof}[Proof of \cref{lem:multivariate-smoothing}]
To establish the privacy ratio, fix an outcome $y \in \mathcal{Y}^T$, a coordinate $j \in [d]$, and a vector $v \in \R^d$.
Conditioning on the noise vector $Z_{-j} = z_{-j} \in \R^{d-1}$ for all coordinates other than $j$, the 1D slice function $f_{z_{-j}}(x) \coloneqq p_{(v_1+z_1, \ldots, x, \ldots, v_d+z_d)}(y)$ takes values in $(0, 1]$ and satisfies the curvature hypothesis $\frac{d^2}{dx^2}\log f_{z_{-j}}(x) \geq -r^2$.
The conditional expectation over $Z_j$, defined by $H_{z_{-j}}(x) \coloneqq \E_{Z_j}[f_{z_{-j}}(x + Z_j)]$, corresponds precisely to the mixture in \cref{prop:scalar-smoothing}.
Applying \eqref{eq:prop-shift-bound} yields $H_{z_{-j}}(v_j + t) \leq e^{L\abs{t}} H_{z_{-j}}(v_j)$ for all $t \in \R$.
Integrating this bound over the distribution of $Z_{-j}$ demonstrates that shifting a single coordinate preserves the multiplicative bound:
\begin{equation}\label{eq:coordinate-shift}
    Q_{v + t e_j}(y) = \E_{Z_{-j}}\left[ H_{Z_{-j}}(v_j + t) \right] \leq e^{L \abs{t}} \E_{Z_{-j}}\left[ H_{Z_{-j}}(v_j) \right] = e^{L \abs{t}} Q_v(y).
\end{equation}

For arbitrary inputs $v, v' \in \R^d$, we interpolate from $v$ to $v'$ along the standard coordinate-wise hybrid chain $v^{(k)} \coloneqq (v'_1, \ldots, v'_k, v_{k+1}, \ldots, v_d)$, with $v^{(0)} = v$ and $v^{(d)} = v'$.
Applying \eqref{eq:coordinate-shift} at each step gives $Q_{v^{(k)}}(y) \leq e^{L\abs{v'_k - v_k}} Q_{v^{(k-1)}}(y)$.
Multiplying these bounds across all $k \in [d]$ yields the pointwise inequality
\begin{equation*}
    Q_{v'}(y) \leq \left( \prod_{k=1}^d e^{L\abs{v'_k - v_k}} \right) Q_v(y) = e^{L \norm{v' - v}_1} Q_v(y).
\end{equation*}
Summing over all atoms $y \in F$ extends this guarantee to any event $F \subseteq \mathcal{Y}^T$, establishing \eqref{eq:event-privacy}.

It remains to analyze the proximity to the unperturbed distribution.
Let $E_0 \coloneqq \bigcap_{j=1}^d \{Z_j = 0\}$ denote the event that no coordinate receives Laplace perturbation.
By coordinate independence, $\Prb(E_0) = (1-\pi)^d$.
Because $p_{v+z}(y) \geq 0$ everywhere, conditioning on $E_0$ and dropping the non-negative contribution on $E_0^c$ gives
\begin{equation*}
    Q_v(y) = \E[p_{v+Z}(y)] \geq \Prb(E_0) \E[p_{v+Z}(y) \mid E_0] = (1-\pi)^d p_v(y),
\end{equation*}
which proves the pointwise domination bound.
Furthermore, we may couple the draws from $Q_v$ and $p_v$ such that they coincide whenever $E_0$ occurs.
By \cref{lem:coupling}, the total variation distance between the two laws is therefore bounded by the probability of the failure event: $\norm{Q_v - p_v}_{\TV} \leq \Prb(E_0^c) = 1 - (1-\pi)^d \leq d\pi$.
\end{proof}
Again, \cref{lem:multivariate-smoothing} does not say that a rare Laplace perturbation privatizes an arbitrary mechanism. It requires a hypothesis on a global, pointwise lower bound on the log curvature. In the following sections we will verify that hypothesis for an entire adaptive query-selection transcript.

\begin{remark}[Connection to Pure Differential Privacy]
In the context of statistical query release on a database histogram $h \in \N^d$ of size $N$, replacing a single individual's record changes the histogram to an adjacent $h'$ with $\norm{h' - h}_1 \leq 2$.
Applying \cref{lem:multivariate-smoothing} with parameters $\kappa = \eps / 8$ and $r = \lambda \sqrt{T} \leq \eps / (8u)$ guarantees an overall privacy parameter of
\begin{equation*}
    L \norm{h' - h}_1 \leq 2 L = 2(\kappa + ru) \leq 2\left(\frac{\eps}{8} + \frac{\eps}{8}\right) = \frac{\eps}{2},
\end{equation*}
achieving pure differential privacy on the smoothed mixture.
Meanwhile, setting $\pi = \Theta(\beta / d)$ ensures that $\norm{Q_h - p_h}_{\TV} \leq d\pi = O(\beta)$, meaning the algorithm executes on the exact unperturbed histogram $h$ with probability at least $1 - O(\beta)$.
\end{remark}

\section{A Selection-Only Multiplicative Weights Process}
\label{sec:transcript}
Fix a public query workload $\mathcal{Q} = \{q_1, \cdots ,q_k\}$. Set \(K=2k\) and construct the signed list
\(s_{2i-1}=q_i\), \(s_{2i}=-q_i\) for every \(i\in[k]\).
Labels remain distinct even when two signed vectors agree.
For every \(H \in \mathbb{R}^d_{\geq 0}\) of mass \(N\), we define
\begin{equation}\label{eq:signed-max}
 \max_{i\in[K]}\ip{H-h}{s_i}
       =N\err_{\cQ}(H/N,h).
\end{equation}
As illustrated in \cref{sec:overview-framework}, we use the following algorithm for privately updating the candidate histogram. Note that though the input histogram $h\in \mathbb{N}^d$, \cref{alg:base} is well-defined for every real-valued input $v\in \mathbb{R}^d$.
\begin{algorithm}[h]
\caption{Selection-only process \(\mathsf{SelectMW}(v)\)}
\label{alg:base}
\begin{algorithmic}[1]
\Require \(v\in\R^d\); public \(N,T\in \mathbb{N}_{\geq 1}, \lambda>0\),
         \(0<\rho\leq1/2\), and signed list \(s_1,\ldots,s_K\).
\State Initialize \(p_1 = (1/d, \ldots, 1/d) \in \Delta_d\).
\For{\(t=1,\ldots,T\)}
 \State Set \(H_t=Np_t\).
 \State Draw \(i_t\in[K]\) with probabilities
 \[
   P_v(i_t=i\mid i_1,\ldots,i_{t-1})
   =\frac{\exp(\lambda\ip{H_t-v}{s_i})}
          {\sum_{a=1}^K\exp(\lambda\ip{H_t-v}{s_a})}.
 \]
 \If{\(t < T\)}
   \State Update for each \(j \in [d]\):
   \[
     p_{t+1}(j) = \frac{p_t(j)(1-\rho s_{i_t}(j))}{\sum_{a=1}^d p_t(a)(1-\rho s_{i_t}(a))}.
   \]
 \EndIf
\EndFor
\State Return \(y=(i_1,\ldots,i_T)\).
\end{algorithmic}
\end{algorithm}
Let $y = (i_1, \cdots, i_{T})$ be the transcript of the private selection of \emph{bad} queries. Clearly, since each update of $p_t$ in \cref{alg:base} is decided by $i_t$, there is a deterministic decoder that maps a transcript to
\begin{equation}\label{eq:decoder}
 D(y)=\frac1T\sum_{t=1}^T p_t\in\Delta_d.
\end{equation}
In particular, the decoder is a function only of the transcript and the public inputs, even when \(v\) is not a nonnegative histogram.
Write \(p_v(y)\) for the transcript probability in
\cref{alg:base}.

\subsection{Pointwise Derivatives of the Transcript Law}

\begin{lemma}[Curvature and first-derivative bounds]
\label{lem:curvature}
For every \(v\in\R^d\), \(j\in[d]\), and \(y\in[K]^T\),
the probability \(p_v(y)\) is positive, infinitely differentiable, and
\begin{equation}\label{eq:transcript-derivatives}
 \left|\partial_{v_j}\log p_v(y)\right|\leq2\lambda T,
 \qquad
 \partial_{v_j}^2\log p_v(y)\geq-T\lambda^2.
\end{equation}
\end{lemma}
\begin{proof}
Fix an arbitrary transcript $y = (i_1, \ldots, i_T) \in [K]^T$, a coordinate $j \in [d]$, and an input vector $v \in \R^d$. By the chain rule of probability, the transcript probability factors as
\[
  p_v(y) = \prod_{t=1}^T P_v(i_t \mid i_1, \ldots, i_{t-1}).
\]
And thus,
\begin{equation}\label{eq:log-pv-sum}
  \log p_v(y) = \sum_{t=1}^T \log P_v(i_t \mid i_{<t}),
\end{equation}
where we abbreviate the prefix history by $i_{<t} \coloneqq (i_1, \ldots, i_{t-1})$. Note that in \cref{alg:base}, the sequence of candidate distributions $p_1, \ldots, p_T$ (and hence the candidate histograms $H_t = N p_t$) is determined entirely by the fixed sequence of signed query indices $i_1, \ldots, i_{T-1}$ and the public initial distribution $p_1 = (1/d, \ldots, 1/d)$. Consequently, for a fixed transcript $y$, $\partial_{v_j} H_t = 0$ for all $t \in [T]$ and $j \in [d]$.  For brevity, let $P_t(a) \coloneqq P_v(i_t = a \mid i_{<t})$ denote the conditional probability of selecting index $a \in [K]$. By the definition of the softmax rule in \cref{alg:base},
\[
  P_t(a) = \frac{\exp(\lambda \ip{H_t - v}{s_a})}{Z_t(v)}, \quad \text{where } Z_t(v) \coloneqq \sum_{a=1}^K \exp(\lambda \ip{H_t - v}{s_a}).
\]
Since the exponential function is everywhere strictly positive and smooth on $\R$, each term in the sum is strictly positive, ensuring that $Z_t(v) > 0$ and $P_t(i_t) \in (0, 1)$ is smooth ($C^\infty$) on $\R^d$. Taking the logarithm of $P_t(i_t)$ gives
\begin{equation}\label{eq:single-round-log}
  \log P_t(i_t) = \lambda \ip{H_t - v}{s_{i_t}} - \log Z_t(v).
\end{equation}
We first differentiate \eqref{eq:single-round-log} with respect to $v_j$. Since $H_t$ is independent of $v$, the first term satisfies $\partial_{v_j}(\lambda \ip{H_t - v}{s_{i_t}}) = -\lambda s_{i_t}(j)$. For the log-partition function $\log Z_t(v)$, applying the chain rule:
\begin{align*}
  \partial_{v_j} \log Z_t(v) 
  &= \frac{1}{Z_t(v)} \sum_{a=1}^K \partial_{v_j} \exp(\lambda \ip{H_t - v}{s_a})= \frac{1}{Z_t(v)} \sum_{a=1}^K (-\lambda s_a(j)) \exp(\lambda \ip{H_t - v}{s_a}) \\
  &= -\lambda \sum_{a=1}^K P_t(a) s_a(j)\\
  &= -\lambda \E_{a \sim P_t}[s_a(j)].
\end{align*}
Substituting this back into the derivative of \eqref{eq:single-round-log}, we obtain
\begin{equation}\label{eq:single-round-first-deriv}
  \partial_{v_j} \log P_t(i_t) = -\lambda s_{i_t}(j) + \lambda \E_{a \sim P_t}[s_a(j)].
\end{equation}
Because every signed query function $s_a(j)$ is bounded in $[-1, 1]$ for all $a \in [K]$ and $j\in[d]$, and by the triangle inequality,
\[
  \abs{\partial_{v_j} \log P_t(i_t)} \le \lambda \abs{s_{i_t}(j)} + \lambda \abs{\E_{a \sim P_t}[s_a(j)]} \le  2\lambda.
\]
Summing over all $T$ rounds according to \eqref{eq:log-pv-sum} proves the first-derivative bound:
\[
  \abs{\partial_{v_j} \log p_v(y)} \le \sum_{t=1}^T \abs{\partial_{v_j} \log P_t(i_t)} \le 2\lambda T.
\]
For the second derivative, we have
\begin{equation}\label{eq:second-deriv-expansion}
  \partial_{v_j}^2 \log P_t(i_t) = \partial_{v_j} \left( \lambda \sum_{a=1}^K P_t(a) s_a(j) \right) = \lambda \sum_{a=1}^K s_a(j) \partial_{v_j} P_t(a).
\end{equation}
Using the logarithmic derivative identity $\partial_{v_j} P_t(a) = P_t(a) \partial_{v_j} \log P_t(a)$, and applying \eqref{eq:single-round-first-deriv} to the outcome $a$, we have
\[
  \partial_{v_j} P_t(a) = P_t(a) \left( -\lambda s_a(j) + \lambda \E_{b \sim P_t}[s_b(j)] \right).
\]
Substituting this expression into \eqref{eq:second-deriv-expansion} yields
\begin{align*}
  \partial_{v_j}^2 \log P_t(i_t) 
  &= \lambda \sum_{a=1}^K s_a(j) P_t(a) \left( -\lambda s_a(j) + \lambda \E_{b \sim P_t}[s_b(j)] \right) \\
  &= -\lambda^2 \sum_{a=1}^K P_t(a) s_a(j)^2 + \lambda^2 \left( \sum_{a=1}^K P_t(a) s_a(j) \right) \left( \E_{b \sim P_t}[s_b(j)] \right) \\
  &= -\lambda^2 \left( \E_{a \sim P_t}[s_a(j)^2] - \big(\E_{a \sim P_t}[s_a(j)]\big)^2 \right) \\
  &= -\lambda^2 \Var_{a \sim P_t}(s_a(j)).
\end{align*}
Since the random variable $s_a(j)$ takes values within $[-1, 1]$, its variance is non-negative and
\[
  0 \le \Var_{a \sim P_t}(s_a(j)) = \E_{a \sim P_t}[s_a(j)^2] - \big(\E_{a \sim P_t}[s_a(j)]\big)^2 \le \E_{a \sim P_t}[s_a(j)^2] \le 1.
\]
Therefore, 
\[
  \partial_{v_j}^2 \log P_t(i_t) = -\lambda^2 \Var_{a \sim P_t}(s_a(j)) \ge -\lambda^2.
\]
Finally, summing this bound across all $T$ rounds via \eqref{eq:log-pv-sum} gives
\[
  \partial_{v_j}^2 \log p_v(y) = \sum_{t=1}^T \partial_{v_j}^2 \log P_t(i_t) = -\lambda^2 \sum_{t=1}^T \Var_{a \sim P_t}(s_a(j)) \ge -T \lambda^2,
\]
which establishes the claimed lower bound and completes the proof.
\end{proof}

\subsection{Progress of Selection-Only MWU}\label{sec:progress-of-mwu}
In this section we show the utility loss in multiplicative weights update defined in \cref{alg:base}. The proof follows the selection-only PMW analysis of Section~3.2 and Appendices~A.2--A.3 in Fitzsimons~\cite{Fitzsimons26}:
combine a multiplicative-weights potential bound with a softmax
selection-gap estimate, then average the synthetic iterates.
We adapt this argument to the linear multiplicative update
\(1-\rho s_{i_t}(j)\) and use a bounded martingale estimate
to obtain a high-probability guarantee.  
Before getting into the proof, we introduce some supporting tools below. 

\begin{lemma}[Selection gap from exponential mechanism]\label{lem:softmax}
Let $a_i\in \mathbb{R}$ for each $i\in [K]$. If \(P(i)=e^{\lambda a_i}/\sum_j e^{\lambda a_j}\) on \([K]\)
and \(\lambda>0\), then
\[
 \max_i a_i-\E_{i\sim P}a_i\leq\frac{\log K}{\lambda}.
\]
\end{lemma}
\begin{proof}
This is a standard utility analysis of the exponential mechanism; we include it here only for the sake of completeness. Let \(Z=\sum_{j=1}^K e^{\lambda a_j}\), so that
\(\log P(i)=\lambda a_i-\log Z\).
Multiplying by \(P(i)\) and summing over \(i\) gives
\[
 \sum_{i=1}^K P(i)\log P(i)
 =\lambda\sum_{i=1}^K P(i)a_i-\log Z.
\]
By \cref{lem:kl}, the left-hand side is at least \(-\log K\).
Also, the term corresponding to a maximizer of \(a_i\) is included
in \(Z\), so \(\log Z\geq\lambda\max_i a_i\).
Consequently,
\[
 \lambda\E_{i\sim P}a_i
 =\log Z+\sum_{i=1}^K P(i)\log P(i)
 \geq\lambda\max_i a_i-\log K.
\]
Rearrange and divide by \(\lambda>0\).
\end{proof}

\begin{lemma}[Deterministic MW progress]\label{lem:regret}
For every transcript $(i_1, \cdots, i_T)$ and every histogram \(h\) of mass \(N\),
\begin{equation}\label{eq:regret}
 \sum_{t=1}^T\ip{H_t-h}{s_{i_t}}
       \leq N\left(\frac{\log d}{\rho}+\rho T\right).
\end{equation}
\end{lemma}
\begin{proof}
Fix any transcript. We define the comparator by the true histogram \(p^*=h/N\in\Delta_d\).
Write \(s_t=s_{i_t}\) and \(c_t=\ip{p_t}{s_t}\).
Since \(p_t\) is a probability vector and \(s_t(j)\in[-1,1]\),
we have \(|c_t|\leq1\).
In \cref{alg:base}, the normalizing denominator of the update satisfies
\[
  \sum_{a=1}^d p_t(a)(1-\rho s_t(a)) = \sum_{a=1}^d p_t(a) - \rho \ip{p_t}{s_t} = 1 - \rho c_t.
\]
Hence, the normalized iterate update is equivalently written as
\[
 p_{t+1}(j) = \frac{p_t(j)(1-\rho s_t(j))}{1-\rho c_t}.
\]
The multiplier \(1-\rho s_t(j)\) and the normalizing factor
\(1-\rho c_t\) both belong to \([1/2,3/2]\), because
\(0<\rho\leq1/2\).  In particular, all coordinates of \(p_t\)
remain positive, and the potential \(\Phi_t=\KL(p^*\|p_t)\) is finite. We use the inequalities
\begin{equation}\label{eq:mw-log-bounds}
 -z-z^2\leq\log(1-z)\leq-z\qquad (|z|\leq1/2).
\end{equation}
The upper bound follows from \(\log u\leq u-1\) with \(u=1-z\).
For the lower bound, let \(f(z)=\log(1-z)+z+z^2\).
Then \(f(0)=0\) and
\(f'(z)=z(1-2z)/(1-z)\), which is nonpositive on \([-1/2,0]\)
and nonnegative on \([0,1/2]\).  Hence \(f(z)\geq f(0)=0\)
throughout this interval.  Substituting the normalized update into the potential difference gives
\begin{align*}
 \Phi_{t+1}-\Phi_t
 &=\sum_{j:p^*(j)>0}p^*(j)\log\frac{p_t(j)}{p_{t+1}(j)}\\
 &=\log(1-\rho c_t)-\sum_jp^*(j)\log(1-\rho s_t(j))\\
 &\leq-\rho c_t+\rho\sum_jp^*(j)s_t(j)
                  +\rho^2\sum_jp^*(j)s_t(j)^2\\
 &\leq-\rho\ip{p_t-p^*}{s_t}+\rho^2.
\end{align*}
The first inequality uses \eqref{eq:mw-log-bounds}, and the second
uses \(s_t(j)^2\leq1\) and \(\sum_jp^*(j)=1\).
Rearranging and summing over \(t=1,\ldots,T\) (defining the auxiliary iterate $p_{T+1}$ at $t=T$ by the same update rule) makes the potentials telescope:
\[
 \rho\sum_{t=1}^T\ip{p_t-p^*}{s_t}
 \leq\Phi_1-\Phi_{T+1}+\rho^2T
 \leq\log d+\rho^2T.
\]
For the last inequality, \cref{lem:kl} gives \(\Phi_{T+1}\geq0\)
and \(\Phi_1\leq\log d\), since \(p_1=\Unif([d])\).
Finally, \(H_t-h=N(p_t-p^*)\); multiplying by \(N/\rho\)
proves \eqref{eq:regret}.  The argument holds for every sequence
of selected labels, independently of how that sequence was generated.
\end{proof}
\begin{lemma}[A bounded martingale estimate]\label{lem:martingale}
Let \((\mathcal F_t)_{t=0}^T\) be a filtration, and let \(M_t\) be
\(\mathcal F_t\)-measurable with
\(\E[M_t\mid\mathcal F_{t-1}]=0\) and \(|M_t|\leq b\) almost surely,
where \(b>0\).
For \(0<\theta<1\), with probability at least \(1-\theta\),
\[
 \sum_{t=1}^T M_t \leq b\sqrt{2T\log(1/\theta)}.
\]
\end{lemma}
\begin{proof}
We first bound the conditional moment-generating function of a single increment $M_t$.
Fix $u > 0$. Since the exponential function $z \mapsto e^{uz}$ is convex, on $[-b, b]$ we have:
\[
 e^{uz} \leq \frac{b+z}{2b}e^{ub} + \frac{b-z}{2b}e^{-ub} \qquad \text{for all } z \in [-b, b].
\]
Substituting $z = M_t$ (which lies in $[-b, b]$ almost surely), taking the conditional expectation with respect to $\mathcal F_{t-1}$, and using the martingale difference property $\E[M_t \mid \mathcal F_{t-1}] = 0$, we obtain
\[
 \E[e^{uM_t} \mid \mathcal F_{t-1}]
 \leq \frac{e^{ub} + e^{-ub}}{2}
 = \cosh(ub) \leq e^{u^2b^2/2}.
\]
For the final inequality, we compare the terms of the Taylor expansions:
\[
 \cosh(ub) = \sum_{j=0}^\infty \frac{(ub)^{2j}}{(2j)!}
 \leq \sum_{j=0}^\infty \frac{(u^2b^2/2)^j}{j!}
 = e^{u^2b^2/2},
\]
which holds term-by-term because $(2j)! = \prod_{k=1}^j (2k)(2k-1) \geq \prod_{k=1}^j (2k) = 2^j j!$. Next, consider the cumulative sum $S_t = \sum_{s=1}^t M_s$, with $S_0 = 0$.
Since $S_{t-1}$ is $\mathcal F_{t-1}$-measurable, then
\[
 \E[e^{uS_t}]
 = \E\!\left[ e^{uS_{t-1}} \E[e^{uM_t} \mid \mathcal F_{t-1}] \right]
 \leq e^{u^2b^2/2} \E[e^{uS_{t-1}}].
\]
Unrolling this recurrence across all $T$ steps from the base case $\E[e^{uS_0}] = e^0 = 1$ yields
\[
 \E[e^{uS_T}] \leq e^{Tu^2b^2/2}.
\]
Now, for any $a > 0$, since $u > 0$, the mapping $x \mapsto e^{ux}$ is strictly increasing. Applying Markov's inequality to the non-negative random variable $e^{uS_T}$ yields
\[
 \Prb\{S_T \geq a\}
 = \Prb\{e^{uS_T} \geq e^{ua}\}
 \leq e^{-ua} \E[e^{uS_T}]
 \leq \exp\!\left( -ua + \frac{Tu^2b^2}{2} \right).
\]
Minimizing this quadratic exponent by choosing $u = a/(Tb^2) > 0$, the bound simplifies to
\[
 \Prb\{S_T \geq a\} \leq \exp\!\left( -\frac{a^2}{2Tb^2} \right).
\]
Setting $a = b\sqrt{2T\log(1/\theta)}$ makes the right-hand side equal to $\theta$, so that
\[
 \Prb\!\left\{ \sum_{t=1}^T M_t \geq b\sqrt{2T\log(1/\theta)} \right\} \leq \theta.
\]
Therefore, we conclude that $\sum_{t=1}^T M_t \leq b\sqrt{2T\log(1/\theta)}$ holds with probability at least $1-\theta$.
\end{proof}

\begin{proposition}[Accuracy of the unperturbed process]
\label{prop:base-utility}
Run \(\mathsf{SelectMW}(h)\) on the true histogram $h$.
For every \(0<\theta<1\), with probability at least \(1-\theta\),
\begin{equation}\label{eq:base-utility}
 \err_{\cQ}(D(y),h)
 \leq\frac{\log d}{\rho T}+\rho
       +\frac{\log K}{N\lambda}
       +4\sqrt{\frac{2\log(1/\theta)}T}.
\end{equation}
\end{proposition}
\begin{proof}
Fix the true input histogram $h \in \mathcal{H}_{N,d}$. Let $\mathcal{F}_t = \sigma(i_1, \ldots, i_t)$ denote the natural filtration generated by the first $t$ selected signed query indices, with $\mathcal{F}_0$ being the trivial $\sigma$-algebra. Because the candidate distribution $p_t$ is determined entirely by the preceding indices $i_1, \ldots, i_{t-1}$, the candidate histogram $H_t = N p_t$ is $\mathcal{F}_{t-1}$-measurable.

At round $t$, define the maximum discrepancy across all signed queries and the discrepancy of the actually selected query by
\[
  G_t \coloneqq \max_{a \in [K]} \ip{H_t - h}{s_a}, \qquad
  X_t \coloneqq \ip{H_t - h}{s_{i_t}}.
\]
Conditioned on $\mathcal{F}_{t-1}$, the error vector $(\ip{H_t - h}{s_a})_{a \in [K]}$ is fixed. In \cref{alg:base}, the index $i_t$ is sampled from the softmax distribution with parameter $\lambda$ over these $K$ scores. Applying the softmax selection gap bound (\cref{lem:softmax}) to this conditional distribution yields
\begin{equation}\label{eq:conditional-progress}
  G_t \leq \E[X_t \mid \mathcal{F}_{t-1}] + \frac{\log K}{\lambda}.
\end{equation}
Define the martingale difference sequence
\[
  M_t \coloneqq \E[X_t \mid \mathcal{F}_{t-1}] - X_t, \qquad t \in [T].
\]
By construction, $M_t$ is $\mathcal{F}_t$-measurable and satisfies $\E[M_t \mid \mathcal{F}_{t-1}] = 0$. Since $H_t = N p_t$ with $p_t \in \Delta_d$ and $h \in \mathcal{H}_{N,d}$, their $\ell_1$-norms satisfy $\norm{H_t}_1 = N$ and $\norm{h}_1 = N$. By the triangle inequality, $\norm{H_t - h}_1 \le \norm{H_t}_1 + \norm{h}_1 = 2N$. Because every signed query satisfies $\norm{s_a}_\infty \le 1$, Hölder's inequality implies that for every possible realization of $i_t$,
\[
  |X_t| = |\ip{H_t - h}{s_{i_t}}| \le \norm{H_t - h}_1 \norm{s_{i_t}}_\infty \le 2N.
\]
Consequently, the conditional expectation also lies in the same interval, $\E[X_t \mid \mathcal{F}_{t-1}] \in [-2N, 2N]$. By the triangle inequality, the increment is almost surely bounded:
\[
  |M_t| \le |\E[X_t \mid \mathcal{F}_{t-1}]| + |X_t| \le 2N + 2N = 4N.
\]
Applying the bounded martingale estimate (\cref{lem:martingale}) with uniform bound $b = 4N$, with probability at least $1 - \theta$,
\begin{equation}\label{eq:martingale-bound-event}
  \sum_{t=1}^T M_t \leq 4N \sqrt{2T \log(1/\theta)}.
\end{equation}

Now, rewrite \eqref{eq:conditional-progress} as $G_t \le X_t + M_t + \frac{\log K}{\lambda}$. Summing over all rounds $t = 1, \ldots, T$ and conditioning on the event where \eqref{eq:martingale-bound-event} holds,
\begin{align*}
  \sum_{t=1}^T G_t 
  &\le \sum_{t=1}^T X_t + \sum_{t=1}^T M_t + \frac{T \log K}{\lambda} \\
  &\le \sum_{t=1}^T \ip{H_t - h}{s_{i_t}} + 4N \sqrt{2T \log(1/\theta)} + \frac{T \log K}{\lambda}.
\end{align*}
Applying the deterministic multiplicative-weights progress bound (\cref{lem:regret}) to $\sum_{t=1}^T \ip{H_t - h}{s_{i_t}}$, we see that
\begin{equation}\label{eq:sum-Gt-bound}
  \sum_{t=1}^T G_t \leq N\left( \frac{\log d}{\rho} + \rho T \right) + \frac{T \log K}{\lambda} + 4N \sqrt{2T \log(1/\theta)}.
\end{equation}

It remains to translate this cumulative discrepancy bound into the worst-case workload error of the decoded distribution $\widehat p = D(y) = \frac{1}{T}\sum_{t=1}^T p_t$. Since $N D(y) = \frac{1}{T}\sum_{t=1}^T H_t$, linearity of the inner product implies that for \emph{any} signed query $s_i \in \mathcal{S}$ ($i \in [K]$),
\[
  \ip{N D(y) - h}{s_i} = \frac{1}{T} \sum_{t=1}^T \ip{H_t - h}{s_i}.
\]
Notice that for each individual round $t$, $\ip{H_t - h}{s_i} \le \max_{a \in [K]} \ip{H_t - h}{s_a} = G_t$ holds pointwise. Averaging over $t$ yields
\[
  \ip{N D(y) - h}{s_i} \le \frac{1}{T} \sum_{t=1}^T G_t \qquad \text{for all } i \in [K].
\]
Because this upper bound holds uniformly for every $i \in [K]$, it holds for their maximum:
\[
  \max_{i \in [K]} \ip{N D(y) - h}{s_i} \le \frac{1}{T} \sum_{t=1}^T G_t.
\]
By the signed representation identity \eqref{eq:signed-max}, the left-hand side is $N \err_{\cQ}(D(y), h)$. Dividing both sides by $N$ and substituting \eqref{eq:sum-Gt-bound}, we conclude that
\begin{align*}
  \err_{\cQ}(D(y), h) 
  &\le \frac{1}{NT} \sum_{t=1}^T G_t \\
  &\le \frac{1}{NT} \left[ N\left( \frac{\log d}{\rho} + \rho T \right) + \frac{T \log K}{\lambda} + 4N \sqrt{2T \log(1/\theta)} \right] \\
  &= \frac{\log d}{\rho T} + \rho + \frac{\log K}{N\lambda} + 4 \sqrt{\frac{2\log(1/\theta)}{T}},
\end{align*}
which establishes \eqref{eq:base-utility} with probability at least $1 - \theta$.
\end{proof}

\section{The Private Query-Release Algorithm}\label{sec:ideal}

We now combine the input-smoothing lemma (Lemma~\ref{lem:multivariate-smoothing}) with the analysis of
\(\mathsf{SelectMW}\) in \cref{sec:progress-of-mwu} to prove \cref{thm:hp}. As our main algorithm, \Cref{alg:release} adds one sparse Laplace perturbation to the input, runs the selection-only process (\cref{alg:base}), and releases the average synthetic histogram. We use the parameter ranges of \cref{thm:hp}
throughout this section.

\begin{algorithm}[htbp]
\caption{Pure-private statistical query release via sparse input smoothing}
\label{alg:release}
\begin{algorithmic}[1]
\Require \(h\in\mathcal H_{N,d}\); public query table $\mathcal{Q} = \{q_i\}_{i\in[k]}$,
         \(0<\eps\leq1\), and \(0<\beta<1/2\).
\State Set \(K=2k\) and form the signed queries
       \(s_{2i-1}=q_i\), \(s_{2i}=-q_i\) for \(i\in[k]\).
\State Set the public parameters
\begin{align}
 \ell&=\log K,&
 \pi&=\frac{\beta}{32d},&
 A&=\log\frac4\pi=\log\frac{128d}{\beta}, \notag\\
 T&=\left\lceil\frac{N\eps}{\ell}\right\rceil,&
 \kappa&=\frac\eps8,&
 \lambda&=\frac{\eps}{8\sqrt{2AT}}.
 \label{eq:parameters}
\end{align}
\State Set the learning rate
\begin{equation}\label{eq:rho}
 \rho=\min\left\{\frac12,\sqrt{\frac AT}\right\}.
\end{equation}
\For{\(j=1,\ldots,d\)}
 \State Independently draw \(B_j\sim\operatorname{Bernoulli}(\pi)\)
        and \(L_j\sim\Lap(\kappa)\); set \(Z_j=B_jL_j\).
\EndFor
\State Draw \(y\gets\mathsf{SelectMW}(h+Z)\) using fresh randomness.
\State Return \(\widehat p=D(y)\in\Delta_d\) from \eqref{eq:decoder}.
\end{algorithmic}
\end{algorithm}

\subsection{Privacy Analysis}
\begin{proposition}[Privacy]
\label{prop:ideal-privacy}
\Cref{alg:release} is pure \(\eps\)-differentially private under record replacement.
In particular, the same privacy guarantee holds for its internal transcript distribution \(Q_h = \E_{Z}[p_{h+Z}]\).
\end{proposition}
\begin{proof}
The hidden perturbation $Z$ in \cref{alg:release} follows the sparse Laplace distribution defined in \cref{proc:smoothed-procedure}, so the induced transcript law is given by the smoothed mixture $Q_h = \E_{Z}[p_{h+Z}]$. We then verify that $p_v(y)$ satisfies the conditions of the sparse smoothing lemma (\cref{lem:multivariate-smoothing}). By \cref{lem:curvature}, for every transcript $y \in [K]^T$, the log-likelihood of $p_v(y)$ satisfies $\partial_{v_j}^2 \log p_v(y) \geq -T \lambda^2$ across all $v \in \R^d$ and $j \in [d]$. We may therefore set the curvature parameter to
\[
  r \coloneqq \lambda \sqrt{T} = \frac{\eps}{8\sqrt{2A}}.
\]
Recall from \eqref{eq:parameters} that $\pi = \beta / (32d)$ and $A = \log(4/\pi) = \log(128d/\beta)$. Since $d \geq 2$ and $\beta < 1/2$, we have $128d/\beta \geq 512$, which implies $A \geq \log 512 > 6$. Consequently, $\sqrt{2A} > 1$, which verifies that
\[
  r = \frac{\eps}{8\sqrt{2A}} < \frac{\eps}{8} = \kappa.
\]
The prerequisite $0 < r \leq \kappa$ of \cref{lem:multivariate-smoothing} is therefore satisfied.

Next, we evaluate the log-Lipschitz constant $L$ from \eqref{eq:L-def}. Substituting $\sqrt{2\log(4/\pi)} = \sqrt{2A}$ and the parameters $\kappa, r$ yields
\[
  L = \kappa + r \sqrt{2\log(4/\pi)} = \frac{\eps}{8} + \left( \frac{\eps}{8\sqrt{2A}} \right) \cdot \sqrt{2A} \leq \frac{\eps}{2}.
\]
Under record replacement, any two neighboring database histograms $h, h' \in \mathcal{H}_{N,d}$ satisfy $h' = h + e_a - e_b$ for some $a, b \in [d]$, which implies $\norm{h' - h}_1 \leq 2$. Applying the pure privacy bound \eqref{eq:event-privacy} of \cref{lem:multivariate-smoothing}, for every measurable transcript event $F \subseteq [K]^T$,
\[
  Q_{h'}(F) \leq e^{L \norm{h' - h}_1} Q_h(F) \leq e^{2L} Q_h(F) \leq e^{\eps} Q_h(F).
\]
This proves that the internal transcript law $Q_h$ satisfies pure $\eps$-differential privacy.

Finally, the released synthetic distribution $\widehat{p} = D(y)$ is a deterministic function of the transcript $y$ and public parameters, so combining the post-processing property of differential privacy (\cref{lem:postprocess}) completes the proof.
\end{proof}
\subsection{Accuracy Analysis}
\begin{proposition}[High-probability accuracy]
\label{prop:ideal-utility}
There is a universal constant \(C_0\) such that, for every
\(h\in\mathcal H_{N,d}\), the output \(\widehat p=D(y)\) of
\cref{alg:release} satisfies, with probability at least
\(1-9\beta/32\),
\[
 \err_{\cQ}(\widehat p,h)\leq
 C_0\min\left\{1,
    \sqrt{\frac{\log(2d/\beta)\log(2k)}{N\eps}}\right\}.
\]
\end{proposition}
\begin{proof}
Let $E_0 \coloneqq \bigcap_{j=1}^d \{Z_j = 0\}$ denote the clean event where no coordinate receives Laplace perturbation in \cref{alg:release}. By a union bound over the $d$ coordinates,
\[
  \Prb(E_0^c) = \Prb\left( \bigcup_{j=1}^d \{B_j = 1\} \right) \le \sum_{j=1}^d \pi = d\pi = \frac{\beta}{32}.
\]
Conditioned on $E_0$, the input histogram is completely unperturbed ($h + Z = h$), and the mechanism executes the exact unperturbed base process $\mathsf{SelectMW}(h)$ (\cref{alg:base}) with transcript law $p_h$.

We now condition on $E_0$ and apply the utility guarantee (\cref{prop:base-utility}) on $p_h(y)$ with failure parameter $\theta = \beta/4$. By a union bound, \cref{alg:release} satisfies \eqref{eq:base-utility} with probability at least
\[
  1 - \Prb(E_0^c) - \theta \ge 1 - \frac{\beta}{32} - \frac{\beta}{4} = 1 - \frac{9\beta}{32}.
\]
It remains to simplify the right-hand side of \eqref{eq:base-utility} under our parameter choices.

Recall from \eqref{eq:parameters} that $\ell = \log K = \log(2k)$ and $A = \log(4/\pi) = \log(128d/\beta)$. Since $d \ge 2$ and $\beta < 1/2$, we have $128d/\beta \ge 512$, which implies $A \ge \log 512 > 6$. Moreover:
\[
  \log d \le A, \qquad \log(4/\beta) \le A, \qquad \log K = \ell.
\]
Define the normalized scale $x \coloneqq \frac{N\eps}{\ell} > 0$ and the target ratio $R \coloneqq \sqrt{A/x}$. Since the  queries $q_i\in \mathcal{Q}$ take values in $[0, 1]$ and both $\widehat{p}$ and $h/N$ are probability distributions in $\Delta_d$ ,the worst-case query error always satisfies the trivial deterministic bound $\err_{\cQ}(\widehat p, h) \le 1$. We now divide the analysis into two cases based on the number of rounds $T = \lceil x \rceil$:

\paragraph{Case 1: $T < 4A$.}
In this regime, we have $x \le T < 4A$, which implies
\[
  R = \sqrt{\frac{A}{x}} > \sqrt{\frac{A}{4A}} = \frac{1}{2}.
\]
Consequently, the deterministic upper bound satisfies
\[
  \err_{\cQ}(\widehat p, h) \le 1\le 2 \min\{1, R\}.
\]

\paragraph{Case 2: $T \ge 4A$.}
Because $A > 6$, we have $T \ge 4A > 24$, which ensures $x > T - 1 \ge 24$, and therefore $x \le T \le x + 1 \le 2x$.
Furthermore, $T \ge 4A$ implies $\sqrt{A/T} \le \sqrt{A/(4A)} = 1/2$. Thus, the learning rate in \eqref{eq:rho} becomes $\rho = \sqrt{A/T} \le 1/2$.
We now bound each term appearing on the right-hand side of \eqref{eq:base-utility}:
\begin{enumerate}[label=(\roman*)]
  \item Using $\log d \le A$ and $\rho = \sqrt{A/T}$, the MW optimization terms satisfy:
  \[
    \frac{\log d}{\rho T} + \rho \le \frac{A}{\sqrt{AT}} + \sqrt{\frac{A}{T}} = 2\sqrt{\frac{A}{T}}.
  \]
  \item Recalling $\lambda = \frac{\eps}{8\sqrt{2AT}}$ and $x = \frac{N\eps}{\ell}$, the selection error satisfies:
  \[
    \frac{\log K}{N\lambda} = \frac{\ell}{N} \cdot \frac{8\sqrt{2AT}}{\eps} = \frac{8\sqrt{2AT}}{x}.
  \]
  \item With $\theta = \beta/4$ and $\log(4/\beta) \le A$, the martingale concentration term satisfies:
  \[
    4\sqrt{\frac{2\log(1/\theta)}{T}} = 4\sqrt{\frac{2\log(4/\beta)}{T}} \le 4\sqrt{2}\sqrt{\frac{A}{T}} \le 6\sqrt{\frac{A}{T}}.
  \]
\end{enumerate}
Summing these three contributions, the error bound in \eqref{eq:base-utility} is at most
\[
  2\sqrt{\frac{A}{T}} + 6\sqrt{\frac{A}{T}} + \frac{8\sqrt{2AT}}{x} = 8\sqrt{\frac{A}{T}} + \frac{8\sqrt{2AT}}{x}.
\]
Since $T \ge x$, the first term satisfies $8\sqrt{A/T} \le 8\sqrt{A/x} = 8R$. For the second term, using $T \le 2x$ gives
\[
  \frac{8\sqrt{2AT}}{x} = 8\sqrt{2} \sqrt{\frac{A}{x}} \sqrt{\frac{T}{x}} \le 8\sqrt{2} R \sqrt{2} = 16R.
\]
Combining both terms yields an upper bound of $8R + 16R = 24R$. Combining this with the deterministic bound $\err_{\cQ} \le 1$, we obtain $\err_{\cQ}(\widehat p, h) \le \min\{1, 24R\} \le 24 \min\{1, R\}$.

In both cases, we have established that $\err_{\cQ}(\widehat p, h) \le 24 \min\{1, R\}$. Since $d \ge 2$ and $\beta < 1/2$, we have $2d/\beta \ge 8$, which implies $\log 64 = 2\log 8 \le 2\log(2d/\beta)$, and thus $ A = \log\left(\frac{128d}{\beta}\right) = \log\left(\frac{2d}{\beta}\right) + \log 64\le 3\log(2d/\beta)$.
Substituting $x = \frac{N\eps}{\log(2k)}$ into $R = \sqrt{A/x}$, we obtain
\[
  R \le \sqrt{\frac{3\log(2d/\beta)\log(2k)}{N\eps}} \le \sqrt{3} \sqrt{\frac{\log(2d/\beta)\log(2k)}{N\eps}}.
\]
Setting $C_0 \coloneqq 24\sqrt{3}$ completes the proof.
\end{proof}


\subsection{Proof of the Main Theorem}

\begin{proof}[Proof of \cref{thm:hp}]
By \cref{prop:ideal-privacy}, \cref{alg:release} is pure
\(\eps\)-DP. By \cref{prop:ideal-utility}, the error bound fails with probability
at most \(9\beta/32<\beta\).  This proves the accuracy claim with \(C=24\sqrt{3}\).

For the running time, we count operations in the model of \cref{rem:computation}.
Forming the signed query list takes \(O(dk)\) operations.
The independent Bernoulli and Laplace draws take \(O(d)\) operations
under our unit-cost sampling assumptions.
In each of the \(T\) rounds, computing all \(K=2k\) inner products
costs \(O(dk)\) operations.  Evaluating the softmax weights and
sampling from their cumulative probabilities costs \(O(k)\),
and updating the weights costs \(O(d)\).
The decoder can accumulate the average of the iterates in
\(O(d)\) operations per round.  Since
\(T=\lceil N\eps/\log(2k)\rceil\geq1\), the total is \(O(dkT)\).
Since \(\eps\leq1\) and \(k\geq1\), we have
\(T\leq N/\log2+1=O(N)\), so the running time is polynomial
in \(d,k,N\), as claimed.
\end{proof}

\begin{remark}[Computational model]\label{rem:computation}
We measure running
time in an idealized model that assigns unit cost to each of the
following: arithmetic operations (addition, subtraction,
multiplication, and division) and comparisons on real numbers,
performed without rounding error; exact evaluation of elementary
functions such as exponentials, logarithms, and square roots; and
drawing an independent uniform random variable from \((0,1)\).
These uniform draws supply the randomness for the algorithm's
Bernoulli and softmax samples.  We also assume unit-cost sampling
from the Laplace distribution at any specified positive rate.
\end{remark}
We also directly state the following sample-complexity formulation of our result by inverting the utility guarantee of \cref{thm:hp}.
\begin{corollary}[Sample complexity]\label{cor:sample}
There exists a universal constant $C > 0$ such that the following holds. 
Let $\mathcal{X}$ be a data universe of size $d \geq 2$, and let $\mathcal{Q} = \{q_1, \ldots, q_k\}$ be a workload of $k$ statistical queries over $\mathcal{X}$. For any target accuracy $\alpha \in (0, 1)$, failure probability $\beta \in (0, 1/2)$, and privacy budget $0<\eps\leq 1$, if the database size satisfies
\begin{equation}\label{eq:sample-complexity}
  N \geq C \cdot \frac{\log(2d/\beta) \log(2k)}{\eps \alpha^2},
\end{equation}
then \cref{alg:release} is pure $\eps$-differentially private under record replacement and, given any input database $X \in \mathcal{X}^N$ (with histogram $h \in \mathcal{H}_{N,d}$), outputs a synthetic distribution $\widehat{p} \in \Delta_d$ such that
\begin{equation}\label{eq:sample-utility}
  \Prb\!\left[ \err_{\cQ}(\widehat{p}, h) \le \alpha \right] = \Prb\!\left[ \max_{q \in \mathcal{Q}} \left| \ip{\widehat{p}}{q} - \ip{\frac{h}{N}}{q} \right| \leq \alpha \right] \geq 1 - \beta.
\end{equation}
\end{corollary}
\section*{AI Disclosure}
The idea of using the second derivative of the log-likelihood function to control the overall privacy loss first emerged during interactions with OpenAI's GPT-5.6 Pro on a different problem. We subsequently abstracted the idea and realized that it applies more generally to statistical queries. OpenAI's GPT 6 Astra was prompted with our own idea of applying a sparse mixture of Laplace convolutions to regularize the first-order log-derivatives, assisting with technical derivation. The authors have verified, significantly simplified, and rewritten the algorithms and proofs to improve readability and presentation.
\section*{Acknowledgment}
The authors are grateful to Jingcheng Liu for insightful discussions during the initial stage of this project. In particular, Jingcheng inspired the intuition of using a diffusive perturbation (or convolution smoothing) to exploit the second-derivative lower bound, which directly led to the design of the Laplace convolution framework. This work is supported in part by National Natural Science Foundation of China Grant No. 62472212.
\bibliographystyle{alphaurl}
\bibliography{references}

\clearpage
\appendix
\section{Proof of the Scalar Smoothing}\label{app:scalar-smoothing}

\ScalarSmoothing*

\begin{proof}
We first establish the regularity and intrinsic smoothness of the Laplace convolution $g(x)$.
By performing the change of variables $y = x+t$, the convolution takes the form $g(x) = \int_{-\infty}^\infty \frac{\kappa}{2} e^{-\kappa \abs{x-y}} f(y) \, dy$.
Splitting the domain at $y = x$ eliminates the absolute value in the kernel:
\begin{equation}\label{eq:g-split}
    g(x) = \frac{\kappa}{2} \int_{-\infty}^x e^{-\kappa(x-y)} f(y) \, dy + \frac{\kappa}{2} \int_x^\infty e^{-\kappa(y-x)} f(y) \, dy.
\end{equation}
Because $f$ is continuous and bounded on $\R$, Leibniz's differentiation rule allows differentiation under the integral sign.
Evaluating the derivatives of the variable limits and of the exponential kernels gives
\begin{align*}
    g'(x) &= \frac{\kappa}{2} \left[ e^0 f(x) - \kappa \int_{-\infty}^x e^{-\kappa(x-y)} f(y) \, dy \right] + \frac{\kappa}{2} \left[ -e^0 f(x) + \kappa \int_x^\infty e^{-\kappa(y-x)} f(y) \, dy \right].
\end{align*}
The boundary evaluation terms identically cancel, $\frac{\kappa}{2} f(x) - \frac{\kappa}{2} f(x) = 0$.
Factoring out $\kappa$ from the remaining integrals and applying the triangle inequality relative to \eqref{eq:g-split}, we obtain the uniform gradient bound
\begin{equation}\label{eq:g-lip-bound}
    \abs{g'(x)} \leq \kappa \left[ \frac{\kappa}{2} \int_{-\infty}^x e^{-\kappa(x-y)} f(y) \, dy + \frac{\kappa}{2} \int_x^\infty e^{-\kappa(y-x)} f(y) \, dy \right] = \kappa g(x).
\end{equation}
Consequently, $H(x) = (1-\pi)f(x) + \pi g(x)$ is continuously differentiable, with $$H'(x) = (1-\pi)f'(x) + \pi g'(x).$$

Write $\psi(x) = \log f(x)$. To bound the ratio $\abs{H'(x)} / H(x)$, fix $x \in \R$ and denote the local logarithmic slope of the unperturbed function by $s \coloneqq \abs{\psi'(x)} = \abs{f'(x)} / f(x)$.
When this slope is already moderate, satisfying $s \leq \kappa + ru$, the bound follows directly: using $\abs{f'(x)} = s f(x) \leq (\kappa + ru)f(x)$ together with \eqref{eq:g-lip-bound}, we have
\begin{equation*}
    \abs{H'(x)} \leq (1-\pi) s f(x) + \pi \kappa g(x) \leq (\kappa + ru)\big[(1-\pi)f(x) + \pi g(x)\big] = (\kappa + ru) H(x).
\end{equation*}

The delicate case arises when the unperturbed slope is large, namely $s > \kappa + ru$.
In this regime, we define the normalized excess slope $a \coloneqq (s - \kappa)/r$, which satisfies $a > u > 1$ by construction.
Let $\sigma \coloneqq \operatorname{sign}(\psi'(x)) \in \{-1, +1\}$, so that $\sigma \psi'(x) = s$.
Integrating the curvature lower bound $\psi'' \geq -r^2$ from the base point $x$ along the ray of positive slope, we find that for every $t \geq 0$,
\begin{equation*}
    \sigma \psi'(x + \sigma t) = s + \int_0^t \sigma^2 \psi''(x + \sigma \tau) \, d\tau \geq s - r^2 t.
\end{equation*}
Integrating this slope bound once more with respect to $t$ yields the quadratic growth lower bound
\begin{equation*}
    \psi(x + \sigma t) - \psi(x) = \int_0^t \sigma \psi'(x + \sigma \tau) \, d\tau \geq s t - \frac{1}{2} r^2 t^2,
\end{equation*}
which upon exponentiation reveals that:
\begin{equation}\label{eq:f-quadratic-lower}
    f(x + \sigma t) \geq f(x) \exp\left( st - \frac{1}{2} r^2 t^2 \right) \quad \text{for all } t \geq 0.
\end{equation}

This quadratic surge in $f$ forces the convolution $g(x)$ to be substantial.
Indeed, restricting the integration domain of $g(x)$ to the ray $\{x + \sigma t : t \geq 0\}$ and substituting \eqref{eq:f-quadratic-lower},
\begin{equation*}
    g(x) \geq \frac{\kappa}{2} \int_0^\infty e^{-\kappa t} f(x + \sigma t) \, dt \geq \frac{\kappa}{2} f(x) \int_0^\infty \exp\left( (s-\kappa) t - \frac{1}{2} r^2 t^2 \right) \, dt.
\end{equation*}
Under the substitution $z = rt$ (where $dt = dz/r$), the exponent completes the square as $(s-\kappa)t - \frac{1}{2}r^2t^2 = az - \frac{1}{2}z^2 = \frac{1}{2}a^2 - \frac{1}{2}(z-a)^2$.
Restricting the integral to the unit-length interval $[a, a+1] \subset [0, \infty)$ and recalling that $\kappa \geq r$,
\begin{equation}\label{eq:g-dominance}
    \frac{g(x)}{f(x)} \geq \frac{\kappa}{2r} e^{a^2 / 2} \int_a^{a+1} e^{-\frac{1}{2}(z-a)^2} \, dz \geq \frac{1}{2} e^{a^2 / 2} \cdot e^{-1/2} > \frac{1}{4} e^{a^2 / 2}.
\end{equation}

This lower bound compensates for the small probability $\pi$.
Recalling from \eqref{eq:prop-constants} that $e^{-u^2/2} = \pi/4$, multiplying \eqref{eq:g-dominance} by $\pi$ gives
\begin{equation*}
    \frac{\pi g(x)}{f(x)} \geq \frac{\pi}{4} e^{a^2 / 2} = \exp\left( \frac{a^2 - u^2}{2} \right).
\end{equation*}
Since $a > u > 1$, the function $z \mapsto z$ dominates $z \mapsto 1/z$ pointwise on $[u, a]$.
Integrating this comparison across $[u, a]$ yields $\frac{a^2 - u^2}{2} = \int_u^a z \, dz \geq \int_u^a \frac{1}{z} \, dz = \log(a/u)$, which implies $e^{(a^2-u^2)/2} \geq a/u$.
Consequently, $\frac{\pi g(x)}{f(x)} \geq \frac{a}{u}$, or equivalently $r a f(x) \leq r u \cdot \pi g(x)$.

We now assemble the estimate on $\abs{H'(x)}$ in this steep-slope regime.
Using the identity $s = \kappa + ra$ together with $\abs{g'(x)} \leq \kappa g(x)$,
\begin{align*}
    \abs{H'(x)} &\leq (1-\pi) (\kappa + ra) f(x) + \pi \kappa g(x) \\
                 &= \kappa H(x) + (1-\pi) ra f(x) \\
                 &\leq \kappa H(x) + ra f(x) \\
                 &\leq \kappa H(x) + ru \cdot \pi g(x) \\
                 &\leq (\kappa + ru) H(x),
\end{align*}
where the final inequality follows from $\pi g(x) \leq H(x)$.
This establishes $\abs{H'(x)/H(x)} \leq L$ in all cases.
Finally, integrating $\frac{d}{d\tau} \log H(x + \tau) = \frac{H'(x+\tau)}{H(x+\tau)}$ along the straight line segment between $0$ and $t$ yields $\abs{\log H(x+t) - \log H(x)} \leq L \abs{t}$, and exponentiation completes the proof.
\end{proof}
\end{document}